\documentclass[fleqn,usenatbib,useAMS]{mnras}

\usepackage{xcolor}
\usepackage[normalem]{ulem}

\usepackage{graphicx}	
\usepackage{amsmath}	
\usepackage{amssymb}	
\usepackage{multicol}        
\usepackage{bm}		
\usepackage{pdflscape}	

\usepackage[T1]{fontenc}
\usepackage{ae,aecompl}

\usepackage{newtxtext,newtxmath}

\title[Relation between QPO frequencies and corona radii]{Disc corona radii and QPO frequencies in black hole binaries: testing Lense-Thirring precession origin }

\author[A. Kubota and C. Done et al.]{
Aya Kubota$^{1}$\thanks{E-mail: aya@shibaura-it.ac.jp }, 
Chris Done$^{2}$,  
Kazuki Tsurumi$^1$ and Ryuki Mizukawa$^{1,3}$\\
$^{1}$Department of Electronic Information Systems, Shibaura Institute of Technology, 307 Fukasaku, Minuma-ku,   Saitama-shi, Saitama 337-8570, Japan\\
$^{2}$Department of Physics, University of Durham, South Road, Durham, DH1 3LE,UK\\
$^{3}$Department of Physics, Saitama University, 255 Shimo-ookubo, Sakura-ku, Saitama-shi, Saitama 338-8570, Japan
}
\date{Accepted 2024 January 3. Received YYY; in original form ZZZ} 
\pubyear{2024}
\begin{document}
\label{firstpage}
\pagerange{\pageref{firstpage}--\pageref{lastpage}}
\maketitle

\begin{abstract}

Stellar-mass black hole binary systems in the luminous X-ray states show a strong quasi-periodic oscillation (QPO) in their Comptonised emission. The frequency of this feature correlates with the ratio of a disc to Comptonised emission rather than with total luminosity. Hence it changes dramatically during spectral transitions between the hard and soft states. Its amplitude is also strongest in these intermediate states, making them an important test of QPO models. However, these have complex spectra which generally require a disc and two separate Comptonisation components, making it difficult to uniquely derive the 
spectral parameters. We build a new energy-conserving model of the accretion flow, {\sc SSsed} model, which assumes a fixed radial emissivity but with a changing emission mechanism. This is similar to the {\sc agnsed} model in {\sc xspec} but tuned to be more suitable for stellar mass black holes. It uses a combination of the disc luminosity and temperature to constrain the inner radius of the (colour temperature corrected) blackbody disc, separating this from the more complex Comptonisation spectra emitted inwards of this radius. We show a pilot study of this model fit to hundreds of RXTE spectra of the black hole binary 
XTE~J$1550-564$. We show that the derived disc radius tightly anti-correlates with the central frequencies of the low-frequency QPO detected in the same observations. The relation is consistent with the quantitative predictions of Lense-Thirring precession of the entire inner Comptonisation regions for the assumed system parameters. This supports the scenario that low-frequency QPOs are caused by Lense-Thirring precession.

\end{abstract}
\begin{keywords}
accretion, accretion discs --- X-rays: binaries --- stars: individual (XTE~J$1550-564$)
\end{keywords}


\section{Introduction}
Stellar black hole binaries (hereafter BHBs) are characterised by large 
intensity variation and various spectral states.
The spectral states are classified  mainly based on the ratio of the soft spectral component to the hard spectral component. 
The soft component is well understood as multi-temperature blackbody emission from the optically thick accretion disc,  
while the hard component is considered to be the result of inverse Compton scattering of seed blackbody photons from the disc by high-energy thermal or non-thermal electrons.
The soft state is characterised by a dominant thermal disc component below several keV 
seen together with a weak power-law-like tail of photon index $\Gamma\sim 2.0-2.2$ which extends beyond 500~keV \citep{grove1998,gier1999}. By contrast, the hard state is instead dominated by a hard ($\Gamma\sim 1.4$--1.7) power-law with 
a thermal cutoff at around a few tens of keV to hundreds of keV.
Therefore, there are at least two high-energy electron populations that can provide hard X-ray emission: one consists of steep non-thermal electrons, and the other consists of hot (approximately several tens of keV) thermal electrons.
The former is most clearly important in the soft state, while the latter dominates the hard state.

The spectra in the intermediate states show a more complex structure.
The intermediate states  (further subdivided into the soft and hard intermediate states e.g. \citealt{belloni2005,remillard2006}) 
have spectra in which both the disc and Compton tail are seen, but the tail is much stronger than in the soft state but much softer ($\Gamma > 2.2$) than in the hard state (e.g. \citealt{zycki2001,zdziarski2001,gier2003}).
They can be fit either by two 
different Comptonising plasmsas, or a single plasma which has a hybrid (thermal plus non-thermal) electron distribution such as the {\sc eqpair} model\footnote{https://heasarc.gsfc.nasa.gov/xanadu/xspec/models/eqpair.html} by Paolo Coppi (e.g. \citealt{hja2016})

The disc inner radius can be derived from fitting the expected sum of blackbody spectral models to the data for the expected disc emissivity using a non-relativistic \citep{ss1973} or relativistic \citep{nt1973} standard disc emissivity. These show that this inner radius is generally constant in soft state spectra, as expected if the disc extends down to the 
innermost stable circular orbit (ISCO)
(e.g. the reviews by \citealt{dgk2007,inoue2022}  and references therein). The soft intermediate state (SIMS) spectra 
are likely consistent with a similar 
inner radius (e.g. \citealt{kubota2001,kubota2004a}). 
In contrast, the question of whether the disc reaches the ISCO or is truncated in the hard intermediate state (HIMS) and hard state remains debated. Energetically, the dominant hard X-ray emission is most easily explained with a composite accretion flow, where the thin disc truncates at some radius larger than the ISCO, transitioning into an inner hot geometrically thick accretion flow. 
However, this is claimed to be inconsistent with the results on the inner radius derived from fitting the best current 
models of the reflected emission from an irradiated accretion disc
(e.g. \citealt{wang-ji2018}).

The fast time variability offers an independent way to characterise the spectral states. 
Low-frequency quasi-periodic oscillations (LF-QPOs) are commonly observed in the Comptonisation component in the HIMS, SIMS and bright hard state (\citealt{ingram2019} and references therein). 
Several models have been suggested to explain the origin of LF-QPOs but the one which is currently most popular is the Lense-Thirring (hereafter LT) precession \citep{stella1998} of the entire hot Comptonisation region inwards of the standard disc truncation
(\citealt{ingram2009}, hereafter IDF09; \citealt{motta2014_1655}). 
This model could be tested by measuring the truncation radius between the disc and corona,
and correlating it with the measured QPO frequency. However, it is  difficult to measure 
the disc-corona truncation radius
from the disc continuum in the states where the QPO is strongest as these typically show strong, complex Comptonisation, making the disc parameters difficult to uniquely determine. 

Here we impose additional constraints from energy conservation in order to derive the disc radius across several hundreds of Rossi X-ray Timing Explorer (RXTE; \citealt{smith1998}) data of a stellar BHB, XTE~J$1550-564$. This is now a standard idea in fitting AGN spectra \citep{optxagnf, agnsed}, but was first actually suggested for these complex intermediate state spectra in BHB \citep{done2006}.
We use this to trace the time evolution of the size of the Comptonising corona for all spectral states, 
and compare it with the LF-QPO frequency seen in the same datasets. 
We show that the transition radius between the disc and Comptonisation regions tightly anti-correlates with the LF-QPO centroid frequency, unlike the scatter (parallel tracks) seen in when plotting the LF-QPO against absolute flux. We show that the transition radius derived from the spectral modelling matches fairly well with the quantitative prediction of 
LT precession of the Comptonising regions.

In section~2, we present a summary of RXTE observations of XTE~J$1550-564$ and 
an overview of its spectral and timing features.
In section~3, we describe the details of the new model.
We analyse the spectral data with the new model, and determine the disc-corona geometry for several spectral states
in section~4.
In section~5, we compare the result of spectral analyses  to LF-QPO frequencies and discuss
the validity of LT precession, and
we summarise the results of our study in section~6.
\section{Overview of spectral and timing properties of XTE~J1550--564 }
\label{sec:overview}

XTE~J$1550-564$ is a well studied transient BHB, with five outbursts which  were extensively observed with RXTE. The system parameters are fairly well known, at $d=4.4^{+0.6}_{-0.4}$~kpc, 
with orbital inclination angle $i=75^\circ\pm4^\circ$ and black hole mass 
$M=9.1\pm 0.6M_\odot$, respectively \citep{orosz2002,orosz2011}. 
The Galactic absorption column 
from X-ray spectral fits of
$N_{\rm H}$=(7.5--10)$\times 10^{21}~{\rm cm^{-2}}$ \citep{tomsick2001,corbel2006, miller2003}
is similar to that expected along the line of sight of $N_{\rm H}=9\times 10^{21}~{\rm cm^{-2}}$\citep{dickey1990}, and makes only a small impact on the spectrum above 3~keV measured by RXTE. 

The first two outbursts showed large intensity variations, exhibiting all the states  
(soft, SIMS, HIMS, and hard state), together with strong LF-QPOs, while the 
latter three outbursts remained in the hard states indicating failed outbursts.
Therefore, we focus on the first two outbursts in this study.

\begin{figure*}
 \begin{center}
\includegraphics[width=16cm]{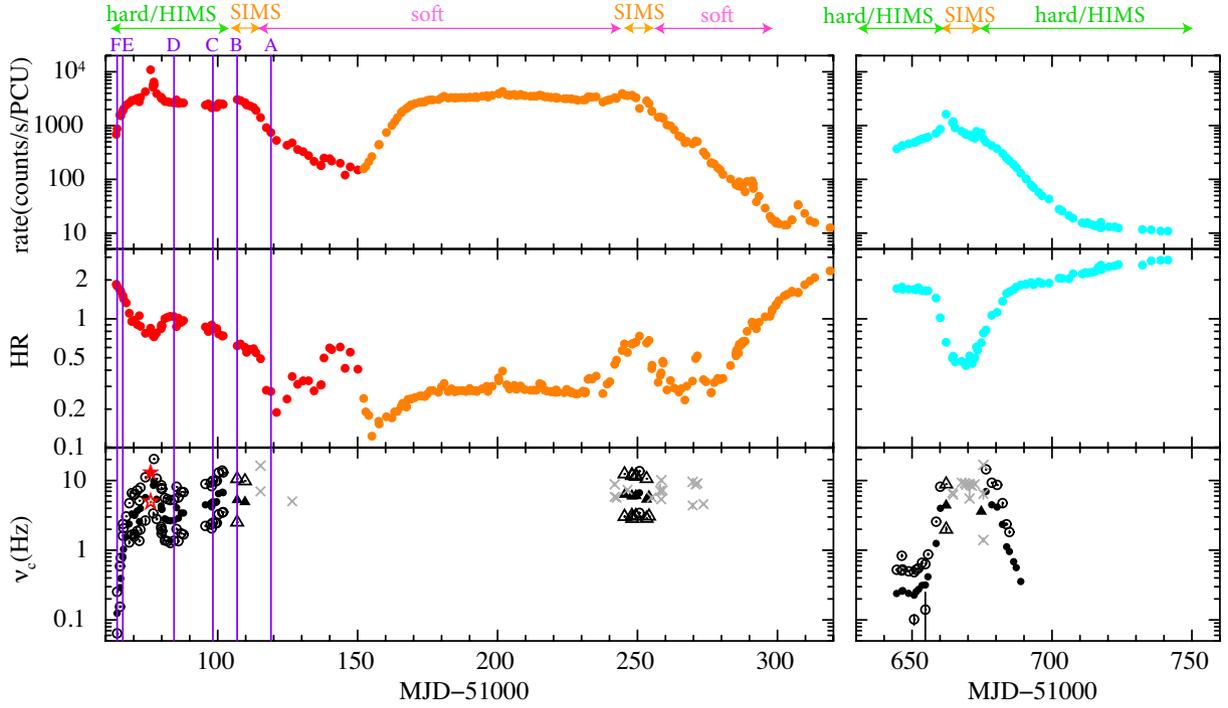} 
 \end{center}
\caption{Light curve of PCA count rate normalised to 1 PCU unit in the 3--20~keV range (top) and evolution of hardness ratio (HR) of the 6--20~keV count rate to the 3--6~keV count rate (2nd panel). The first and second parts of the first outburst are coloured in red and orange, while the second outburst is coloured in cyan.
The lower panel shows the central frequencies of LF-QPOs
in solid symbols, with harmonics/sub-harmonics as open symbols. 
Type-C are denoted by black circles, Type-B by dark grey triangles and Type-A and the other unclear structures with light grey crosses. 
Red stars indicate an irregular type of QPO at which the count rate reached maximum and the strong jet was observed.}
\label{fig:history_12}
\end{figure*}

Figure~\ref{fig:history_12} shows the 3--20~keV PCA light curve normalised to 1 PCU unit (top panel), and the corresponding
time evolution of the hardness ratio (hereafter HR) of 6--20~keV count rate to the 3--6~keV count rate (middle panel).
The first outburst is 
 divided into two semi-outbursts, each of which is indicated by red and orange colours, and the second outburst is colored in cyan. 
The central frequencies of fundamental QPO and its higher/lower harmonics are plotted in the bottom panel of Fig.\ref{fig:history_12}, Type-C and Type-B QPOs marked with black circles and dark grey triangles, respectively,  and Type-A and the other unclear structures  marked with light grey crosses. 
Here, Type-C QPOs are typically observed, and show strong and narrow features with 
broad band, flat-topped noise. Their QPOs are well represented by Lorentzian. 
Type-B QPOs are also narrow and strong but without strong broad band noise, and their structures are reproduced by Gaussian. Type-A QPOs are weak and broad features (\citealt{ingram2019}, and references therein).

Although several authors have already reported the QPO frequencies (e.g. \citealt{remillard2002,rodriguez2004,sobczak2000a}), 
we re-estimated the frequencies using binned mode data to systematically examine their behaviour. 
The centroid frequencies were determined by fitting the power spectral densities (PSDs) with Lorentzians and/or Gaussians.
The calculated frequencies were found to be consistent with the previous results. The LF-QPOs are detected in the SIMS, HIMS and brighter hard state in both rising and decaying phases (${\rm HR}\ge0.5$ and count rates exceeding 100--200~${\rm cts \cdot s^{-1} \cdot PCU^{-1}}$).

\begin{figure*}
\begin{center}
\includegraphics[width=17cm]{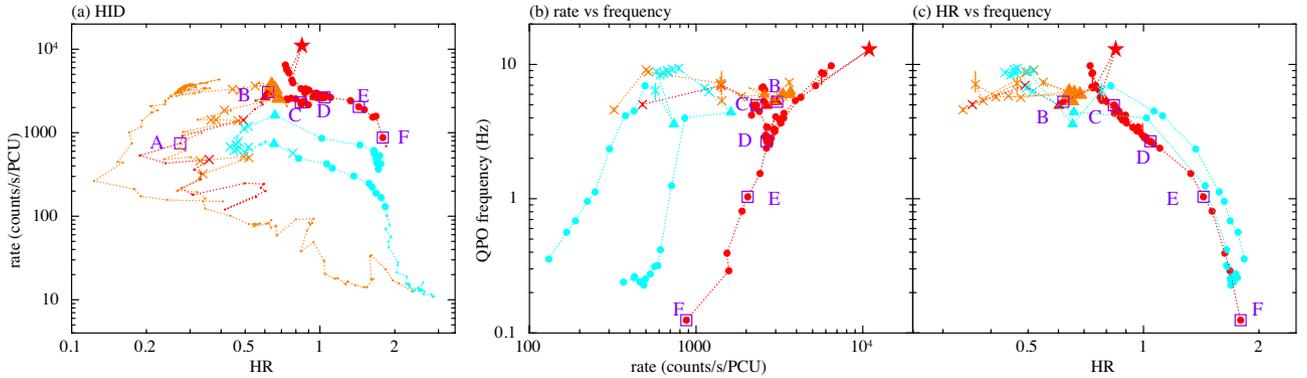}
\end{center}
\caption{Data from Fig.~\ref{fig:history_12} plotted as 
(a) Hardness-Intensity,  (b) Intensity-QPO centroid frequency and (c) Hardness-QPO centroid frequency, also includes the locations of the selected observations A--F 
which characterise the source behaviour in different spectral states and/or different luminosity. Colours indicating the outburst time are the same as in Fig.~\ref{fig:history_12}, as are symbols indicating the QPO types. 
In panel (a), the data points without QPO features are shown as small dots, whereas in (b) and (c) the dots are larger so the points can be seen. 
} 
\label{fig:rate-fc}
\end{figure*}

Figure~\ref{fig:rate-fc}a shows the
hardness-intensity diagram (HID) of these outbursts. 
Data points showing LF-QPOs are identified by different (boldface) symbols, as in
the bottom panel in Fig.~\ref{fig:history_12}, whereas observations without QPOs are marked by small dots. 
We select observations corresponding to characteristic points on this diagram, marked A-F. These are also shown on the time history (Fig.~\ref{fig:history_12}) for reference, and identified in Table~\ref{tab:observation}. 

Figures~\ref{fig:rate-fc}b and \ref{fig:rate-fc}c show the central frequency of the fundamental LF-QPO, $\nu_{\rm c}$, plotted against 
the PCA count rate and HR, respectively. It is clearly evident from Fig.~\ref{fig:rate-fc}b that there are 'parallel tracks', with the same LF-QPO frequency seen at very different 
mass accretion rates (i.e. intensity). Typically the LF-QPOs are seen at higher flux in the fast rise than in the slow decay, and at a different flux level for the fast rise in each outburst. However, these separate tracks merge together when the LF-QPO frequency is plotted instead against HR (Fig.~\ref{fig:rate-fc}c). This shows that the spectral shape (i.e. disc-corona geometry)  
is more important than overall luminosity in determining $\nu_{\rm c}$. 

\begin{table*}
  \caption{Observational data for sample spectra.QPO types were referred from \citet{remillard2002}.
}{%
  \begin{tabular}{cccccccc}\\
      \hline
ID & ObsID & observation start date &MJD &PCA  exposure  &rate  & spectral state & QPO type$^*$\\   
	&	&	& &	(s)	&${\rm  (counts \cdot  s^{-1}\cdot PCU^{-1}})$&	& \\  \hline        
A&30191-01-38-00&1998/11/01 23:56:32&51118&3456&746.6&soft&--\\
B&30191-01-32-00 &1998/10/20 22:52:48& 51106&880&3064.3&SIMS&B\\
C&30191-01-28-02 &1998/10/12 06:33:20 &51098  &1632&2305.5   &HIMS3 &C\\
D&30191-01-14-00&1998/09/28  08:12:48&51084&4704&2644.9&HIMS2&C\\
E&30188-06-01-02&1998/09/10 01:37:20&51066&3424&1885.9&HIMS1&C\\
F&30188-06-03-00&1998/09/08 00:09:52&51064&5664&873.7&hard &C\\
\hline
  \end{tabular}}\label{tab:observation}
\end{table*}

\section{Construction of the SSsed model}

\subsection{Model description}

\subsubsection{Concept of the model}
\label{sub:concept}

As introduced in Section 1, the soft spectral component is generally considered as being produced by the optically thick standard disc, which extends close to the ISCO in the soft state but is truncated far from the ISCO in the hard state. The Comptonisation component requires an electron population which is predominantly thermal in the hard state, 
while non-thermal electrons  are evident in the soft state. This change in accretion flow properties 
makes it difficult to fit the entire outburst with a single model, and transition spectra are especially complicated as they probably require all components so are very difficult to decompose uniquely. 

We build on the model used to constrain similarly complex AGN spectra. The data are often fit with
radially stratified models, 
where an outer blackbody emitting standard disc transitions to a Comptonised disc, where the optically thick material does not quite thermalise but instead is Comptonised in warm plasma to form the soft X-ray excess. The disc truncates in the inner regions, with the flow forming a hot, optically thin plasma close to the black hole (the {\sc agnsed} model in {\sc xspec}: \citealt{agnsed}). This model is (generally) able to uniquely
fit to the complex data even with the absorption gap in the EUV
because the components are energetically constrained by the disc emissivity assuming the mass accretion rate is constant with radius. 

We tailor this approach to be more tuned to BHBs, and especially to their intermediate spectra where there are clearly two Compton components as well as a (truncated) disc. 
The key concept is that the 
flow is radially stratified such that the accretion power is emitted as (colour-corrected) black body radiation at $r > r_{\rm cor}$, while it is emitted as 
inverse-Comptonisation by both the thermal and 
non-thermal corona at $r < r_{\rm cor}$. 
All the emission is constrained by the standard disc emissivity by \cite{ss1973}, with $\dot{M}$ constant with radius. Past work has clearly shown that we can use the temperature and luminosity of the disc component in the high/soft state to determine the mass accretion rate and inner radius of the disc. Here we will use the same idea to derive the geometry of the more complex states, using the shape and luminosity of the low temperature soft component to derive the mass accretion rate through the unComptonised disc and its inner radius in the SIMS, HIMS and hard states, with the remaining radial emissivity used to power the Comptonised emission.

\begin{figure}
 \begin{center}
\includegraphics[width=8cm]{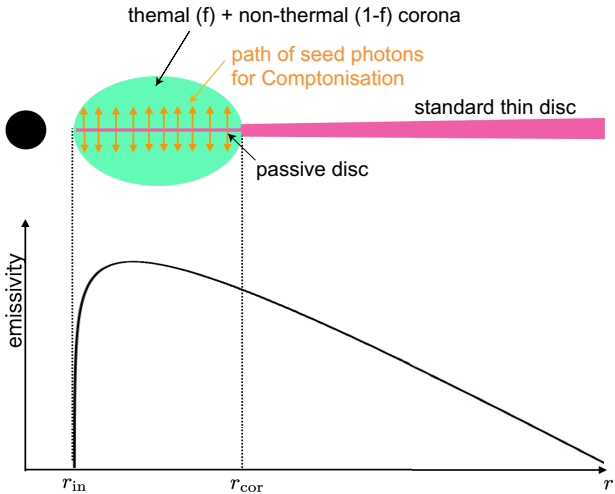} 
 \end{center}
\caption{Sketch of model geometry and radial emissivity profile. 
}
\label{fig:picture}
\end{figure}

\subsubsection{Assumed geometries and emissivity}

The {\sc agnsed} model has a truncated outer blackbody disc which is not colour temperature corrected as this was assumed to be subsumed into the warm Comptonisation region. The higher temperatures of BHB mean that this is not appropriate so we explicitly include the expected colour temperature correction to the blackbody disc emission as cite{optxagnf}. This has assumed angle dependence $\propto \cos i$.

The Comptonising coronal regions of $r_{\rm in}\le r \le r_{\rm cor}$ are less well defined. They can have seed photons from a passive disc below the corona which reprocesses the illuminating flux  (e.g. \citealt{petrucci2013}), or from external direct illumination from the outer disc, or the soft Compton region. External seed photons (from outside rather than from passive disc reprocessing) will add to the emitted luminosity at that radius, making the emissivity at a given radius not exactly defined. 
The angular radiation pattern is also not well defined. If the Comptonisation region is radially extended over the passive disc then it might be 
$\propto \cos i$ or a larger scale height flow might be more isotropic. 

We want to make a model which can be used across all the data, yet
both seed photon source(s) and geometry are probably changing with spectral state. After some experimentation, we make a pragmatic choice to fix the seed photon source as reprocessing from a passive disc underneath the Comptonising coronae, and to assume these emit isotropically. Then the 
local luminosity at radius $r$ 
$F_{\mathrm{cor}}(r) = L_{\mathrm{cor}}(r)/(4\pi d^2)$, while the observed flux from the outer disc ($r>r_{\rm cor}$) is represented as 
$F_{\mathrm{disc}}(r) = L_{\mathrm{disc}}\cos i/(2\pi d^2)$. This geometry of the code is shown in Fig.~\ref{fig:picture}, but we stress that assumption of an underlying passive disc is so as to easily determine a seed photon energy for the model. 

We neglect irradiation of the outer disc by the Comptonisation regions as this has negligible effects on the X-ray spectra of BHBs in the luminous disc states. Reprocessing can affect the  disc emission in the hard state because the intrinsic disc luminosity is small compared to the irradiating Compton luminosity. However, the disc temperature is also low so it is difficult to tightly constrain from RXTE data in this state. 
We also neglect general relativistic ray tracing as we assume these are subsumed into the inclination/geometry uncertainties detailed above. 

The mass accretion rate $\dot{M}$ is assumed to be 
constant throughout the accretion flow from $r_{\rm in}$ to  $r_{\rm out}$, but the efficiency is now defined as in Newtonian gravity as 
$\eta=0.5\cdot(r_{\rm in}/r_{\rm g})^{-1}$. Since $r_{\rm in}$ may now change, 
we normalise the mass accretion rate to the 
Eddington accretion rate defined without including efficiency, so $\dot{m}\equiv \dot{M}/\dot{M}_{\rm Edd}$, where $\dot{M}_{\rm Edd}$ is defined 
as $\dot{M}_{\rm Edd}c^2=L_{\rm Edd}$. 
A changing inner radius could be a real aspect of a large scale height flow, either from magnetic connection allowing it to tap the gravitational potential inwards of the ISCO, or Lense-Thirring torques which truncate the flow above the ISCO.
Alternatively it could just be a small compensation for some of the geometry uncertainties (inclination dependence and relativistic ray tracing) noted above. We term this model {\sc SSsed} (meaning Shakura-Sunyaev SED), with emissivity $\epsilon(r)=3GM \dot{M} f(r)/(8\pi r^3) $ where $f(r)=(1-\sqrt{r_{\rm in}/r})$. 

Details of spectral parameters of the {\sc SSsed} model are shown in Appendix~\ref{app:SSsed_par}.

\subsubsection{Regions of Comptonising components}

It is very clear from the data that there is thermal Comptonisation in the hard state, and non-thermal in the soft state. However, many good quality spectra extending over a broad bandpass require that there are two Compton components, e.g. \cite{yamada2013} and \cite{basak2017} for Cyg X-1, \cite{zdziarski2021} for MAXI~J$1820+070$, 
and \cite{kubota2004b} and \cite{hja2016,connors2019} for XTE~J$1550-564$.

To allow the model to fit across all spectral states we allow the energy to 
be released through hot thermal and/or non-thermal inverse Compton scattering within $r_{\mathrm{cor}}$. We assume the same emissivity as for the standard disc region. This would not be appropriate for the 
advection-dominated accretion flow (ADAF; \citealt{adaf}) in faint states with $\dot{m} \leq 0.1$, where the emissivity is significantly lower than that of the standard disc.
However, in this paper, we focus on the luminous X-ray states where $\dot{m}$ is larger than $0.1$. 

We used the {\sc nthcomp} model \citep{zdziarski1996,zycki1999} to model the Comptonised emission as in {\sc agnsed} model.
The {\sc nthcomp} model describes thermal Comptonisation 
parameterised by photon index $\Gamma$ and electron temperature $kT_{\rm e}$. However, we also use this model  to describe the effect of non-thermal Comptonisation from a steep power law electron distribution. A power law electron distribution produces a power law from inverse Compton scattering, similarly to a Maxwellian (thermal) distribution. The major difference is that thermal electrons produce a cutoff in the spectrum at $\sim 3kT_{\rm e}$, which is not present in the non-thermal case. Instead, a steep electron distribution has a spectral break at 511~keV from the Klein-Nishima cross-section (see e.g. \citealt{hja2016}). 
This can be well approximated by thermal Comptonisation with $kT_{\rm e}=300$~keV in data which do not extend beyond $\sim 150$--200~keV, so we use the {\sc nthcomp} model for this case also.  
Therefore, in this paper, the term 'non-thermal' is employed to signify the absence of a discernible cut-off within the observed energy band.
We allow there to be both thermal and 'non-thermal' electrons in the same region,
parameterised by the 
fraction of accretion power, $f_{\rm th}$, in the thermal Compton, and $\dot{m}_{\rm non}=(1-f_{\rm th})\dot{m}$ in non-thermal Comptonisation. 

We assume the seed photons come from a passive disc underneath the Comptonisation region.
In this case, all the accretion power is supplied to the corona, while the disc itself is originally completely dark. The corona irradiates the disc, which reprocesses the radiation into a blackbody which is as hot as the original standard disc (see e.g. \citealt{agnsed}).
Therefore, the seed photon temperature for the local Comptonisation is set to be the colour-corrected passive disc temperature at radius $r$.
We calculate the released accretion power for a given $\dot{m}$ at radius $r$, and normalise {\sc nthcomp} emission to 
suit for the predicted local luminosity at $r$. This is the same method employed for the disc-corona region in {\sc agnsed} model. We note that the corona intercepts all the photons from the disc, so there is no reflection component from this where the corona is optically thick.

\section{Analyses and Results}
For spectral analyses, we utilised 3--20 keV spectral data from the
PCA and 20--150 keV data from the High-Energy X-ray Timing
Experiment (HEXTE). The source and background spectral files
and the response matrix files, were obtained from the standard data
products provided by the RXTE guest observer facility\footnote{
https://heasarc.gsfc.nasa.gov/docs/xte/recipes/stdprod$\_$guide.html}.
To take systematic uncertainty on calibrations into account\footnote{https://heasarc.gsfc.nasa.gov/docs/xte/pca/doc/rmf/pcarmf-11.7/}, 0.5\% systematic errors were included to each energy bin.

In addition to the continuum emission modelled by the {\sc SSsed}
model, we included a Gaussian line and smeared edge ({\sc smedge}) to
approximately model the reflection features as well as complex 
absorption and emission features around $\sim 6$--10~keV. Since the disc
is expected to be highly ionized, the shape of the reflected emission
is similar to that of the original hard X-ray component, except for
the iron structures around 6--20 keV. Therefore, for efficient computation time 
we opted to use the phenomenological {\sc smedge} and Gaussian models
instead of employing a full reflection code for all the spectral data. 

The spectrum is modified by interstellar absorption, modelled using {\sc tbabs} with the abundance given by \citet{wilms2000}, so the total 
{\sc xspec} model is described as 
\[{\sc tbabs*(smedge*SSsed+Gaussian)}\]
where the values of $M$, $N_{\rm H}$,  $d$ and $i$ are fixed to be $9M_\odot$, $9.0\times 10^{21}~{\rm cm^{-2}}$,
4.4~kpc and $75^\circ$, respectively. 
We also fixed the width of {\sc smedge} at 5~keV. We constrain the edge energy to be 7.2--9.2~keV, central energy $E_{\rm c}$ and $\sigma$ of Gaussian 
to be 6.2--6.9~keV and 0.1--0.8~keV, respectively, and photon indices of two Comptonising components in {\sc SSsed} to be 1.4--4.0.
The normalisations between PCA and HEXTE were adjusted by adding {\sc constant} parameter to the model description, and the presented 
parameters are based on 
the PCA normalisation.

We examine the effect of reflection in more detail in a series of 
Appendices. Firstly we demonstrate that simpler disk plus single Compton and its full reflection cannot fit all the spectra (see Appendix \ref{app:diskbb-nthc}). We then demonstrate that our {\sc SSsed} geometry is not significantly changed from including full reflection on both Compton components compared to the much simpler and faster phenomenological approach used here (see Appendix \ref{app:SSsed-ref}).

\subsection{Example spectral fits to each state}

We first demonstrate that the model can fit all the separate spectral states seen during the outburst, and that the key parameters which are $r_{\rm cor}$ and $\dot{m}$, are not much changed by more detailed reflection modelling. 

\begin{figure*}
 \begin{center}
\includegraphics[width=17cm]{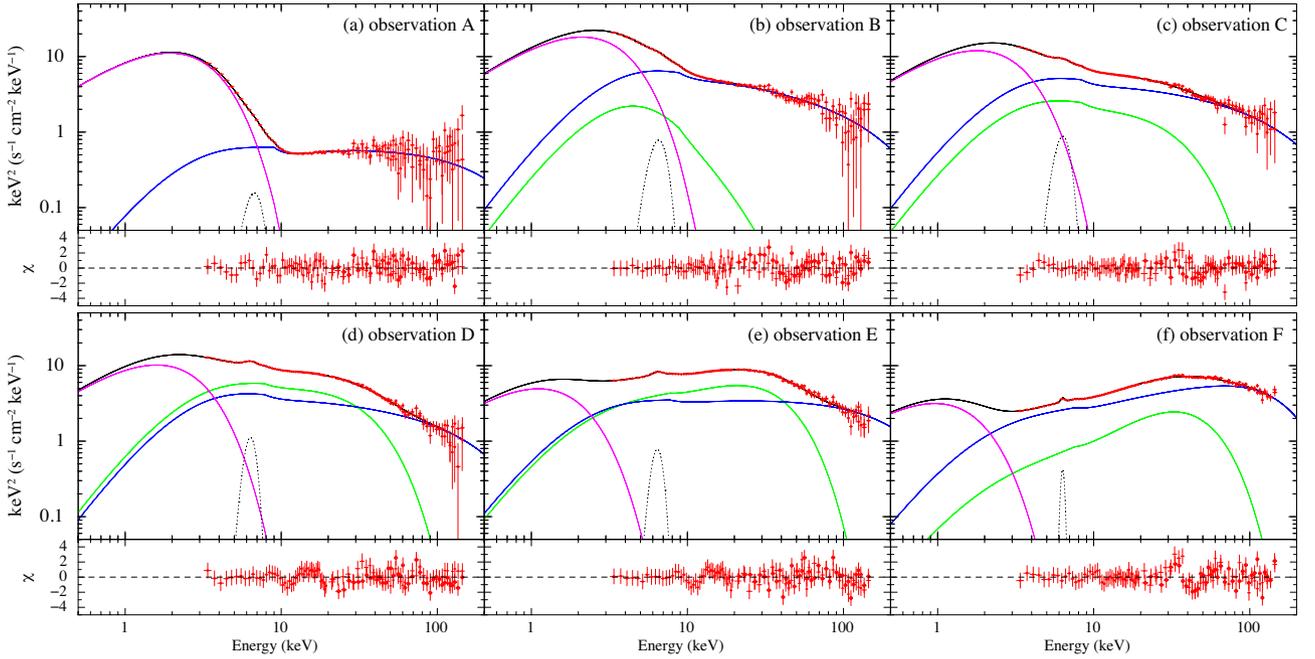}
\end{center}
\caption{Unabsorbed, deconvolved spectra of observations A--F with the best-fitting {\sc SSsed} model and residuals between the data and models. 
Model component of the thermal and non-thermal Comptonising corona regions are shown in green and blue, respectively, 
and that of the outer standard disc is in magenta. 
In observation F, the higher and lower temperature thermal Comptonising components are shown in blue and green, respectively.}
\label{fig:spec_SSsed}
\end{figure*}

Figures~\ref{fig:spec_SSsed}a--\ref{fig:spec_SSsed}f show the spectra of observations A--F 
(identified in Fig.~\ref{fig:rate-fc}a)
with the best-fitting {\sc SSsed} model, with parameters 
shown in Table~\ref{tab:spec_SSsed}. The model components are colour coded with
magenta for the outer standard disc, blue for the non-thermal steep
Compton component, and green for thermal Compton component.
Considering the HR values and count rate, as well as the 
PSDs (shown in Fig.~\ref{fig:psd} in Appendix~\ref{app:psd}), 
the spectral
states of observations A, B, C, D, E and F are consistent with the
soft state, SIMS, (softer)HIMS, HIMS, (harder)HIMS and bright
hard state (as in previous spectral analyses e.g. \cite{kubota2004a,kubota2004b}).

Following the typical spectral studies of the soft state spectra, the
fitting of the soft state data did not include the thermal Comptonising
component (Fig. \ref{fig:spec_SSsed}a). In the SIMS (Figs.~\ref{fig:spec_SSsed}b), it generally also requires a very steep power-law like
component ($\Gamma = 2.5$--4, with an upper limit of $4$) 
together with the non-thermal Comptonising corona ($kT_{\rm e}= 300$~keV
and $\Gamma\ge 2$). 
The HIMS (Figs.~\ref{fig:spec_SSsed}c--\ref{fig:spec_SSsed}e),  most clearly need a combination of thermal and non-thermal Comptonisation to fit the data. The bright hard state (Fig.~\ref{fig:spec_SSsed}f),  is difficult to fit with the thermal and 'non-thermal' approach due to the shape of the cutoff. However, this can work with similar luminosities derived for the total Comptonised emission with more detailed reflection modelling (Appendix~\ref{app:SSsed-ref}). 
This does not change the derived radii and luminosity significantly (see Appendix \ref{app:SSsed-ref}).

We now discuss in more detail the fits to each state. 

\begin{table*}
    \caption{The best-fitting parameters of {\sc SSsed} model for observations A--F.
 The values of $M$, $N_{\rm H}$, $d$, and $i$ are fixed to be $9M_\odot$, $9.0\times 10^{21}~{\rm cm^{-2}}$, 4.4~kpc and $75^\circ$, respectively.  The width of the {\sc smedge} is fixed to be 5~keV. $^\dagger$Values in the parentheses are fixed. $^*$Parameters hit the lower or upper limit. }
  \begin{tabular}{p{2cm}p{2cm}p{2cm}p{2cm}p{2cm}p{2cm}p{2cm}}
      \hline
Obs.             &A&B & C& D & E &F\\
state        &soft&  SIMS &HIMS3& HIMS2 & HIMS1 & hard\\
\hline
\multicolumn{7}{l}{parameters of the {\sc SSsed}}\\
$\log\dot{m}$   &$0.167^{+0.021}_{-0.019}$  &$0.38^{+0.05}_{-0.04}$ &$0.262^{+0.005}_{-0.006}$ & $0.260^{+0.006}_{-0.005}$&$0.115\pm0.009$ & $-0.038^{+0.007}_{-0.006}$\\
  $r_{\rm in}(r_{\rm g})$  &$5.46^{+0.18}_{-0.15}$&$4.4^{+0.4}_{-0.3}$ &(4.5)$^\dagger$&(4.5)$^\dagger$&(4.5)$^\dagger$&(4.5)$^\dagger$\\
   $r_{\rm cor}(r_{\rm g})$   &$7.52^{+0.22}_{-0.20}$&$9.6^{+1.1}_{-0.9}$ &$11.9\pm0.3$ & $15.1^{+0.4}_{-0.6}$   & $25.4^{+1.3}_{-1.1}$ &$28.6^{+1.1}_{-0.9}$\\
$\Gamma_{\rm nth}$ & $2.02\pm0.03$ &$2.34^{+0.03}_{-0.05}$& $2.26^{+0.09}_{-0.10}$  & $2.22^{+0.11}_{-0.16}$ &$1.99^{+0.06}_{-0.08}$&$1.679\pm 0.013$\\
  $kT_{\rm e}$(keV) &(300)$^\dagger$& (300)$^\dagger$& (300)$^\dagger$& (300)$^\dagger$& (300)$^\dagger$&  $46^{+10}_{-5}$\\
 $f_{\rm th}$&(0)$^\dagger$& $0.16\pm0.06$& $0.28\pm 0.13$ & $0.52\pm0.16$ &$0.48 \pm0.09$  & $0.22^{+0.03}_{-0.02}$\\
 $\Gamma_{\rm th}$& --- & $4.0^{+0.0*}_{-0.9}$& $2.23^{+0.08}_{-0.18}$ & $2.08^{+0.07}_{-0.14}$ & $1.70^{+0.07}_{-0.08}$ & $1.40^{+0.05}_{-0.00}$  \\
  $kT_{\rm e}$(keV) &---& (9.0)$^\dagger$  &(9.0)$^\dagger$ & $8.6\pm1.0$ & $8.7^{+0.5}_{-0.6}$ &$10.2^{+0.5}_{-0.4}$\\
\multicolumn{5}{l}{parameters of {\sc smedge} and Gaussian}\\
$E_{\rm edge}$(keV)& $9.01^{+0.19*}_{-0.17}$  &$8.9^{+0.3}_{-0.4}$& $8.3^{+0.4}_{-0.2}$ & $7.82^{+0.26}_{-0.18}$ & $7.76^{+0.21}_{-0.18}$ & $7.57^{+0.14}_{-0.12}$\\
$\tau_{\rm max}$ &$0.89^{+0.11*}_{-0.13}$ &$0.56^{+0.11}_{-0.09}$& $0.52^{+0.10}_{-0.11}$  & $0.47^{+0.08}_{-0.13}$ & $0.34^{+0.06}_{-0.10}$  & $0.42\pm0.04$\\
$E_{\rm c}$(keV)&$6.5\pm0.17$ &$6.36^{+0.20}_{-0.11}$&$6.20^{+0.06}_{-0.00*}$ & $6.27^{+0.09}_{-0.07}$ &$6.34\pm0.08$  & $6.31\pm0.11$\\
$\sigma$(keV)&$0.80^{+0.00*}_{-0.07}$ &$0.77^{+0.03}_{-0.39}$ &$0.66^{+0.14}_{-0.22}$ & $0.51^{+0.25}_{-0.22}$& $0.50^{+0.18}_{-0.17}$ & $0.18^{+0.24}_{-0.08}$\\
norm ($\times 10^{-3}$)& $7.13^{+1.3}_{-1.2}$	 &$36^{+9}_{-18}$& $37 ^{+17}_{-14}$ & $36^{+23}_{-12}$ & $24^{+9}_{-6}$  & $4.7^{+1.5}_{-1.1}$\\
\hline
EW(eV)& $190^{+33}_{-32}$ &$132^{+33}_{-67}$&$165 ^{+72}_{-61}$  & $135^{+89}_{-43}$ & $129^{+46}_{-30}$ & $56^{+19}_{-12}$\\
\hline
$\chi^2$(dof) &109.3(116)&119.2(114)&90.3(115)& 89.2(114) & 113.6(114) & 131.7(113)\\
\hline
  \end{tabular} 
 \label{tab:spec_SSsed}
\end{table*}

\subsection{The soft state and SIMS: inner radius}
\label{sub:soft-sims}

In order to evaluate $r_{\rm in}$,  the {\sc SSsed} model was first applied to the spectra in the soft state. 
As exemplified by observation A shown in Table~\ref{tab:spec_SSsed} and Fig.~\ref{fig:spec_SSsed}a, 
all the data in the soft state were well reproduced with the model of a
disc and a single non-thermal steep Comptonisation
(so we fix $f_{\rm th}=0$)
except for one anomalous spectrum on MJD 51262 (ObsID 40401-01-64-00).

In the soft state, the value of $r_{\rm in}$ were found to be almost constant at (5--$5.5)r_{\rm g}$, which 
corresponds to spin parameter of $a^\ast=0.15$, by considering $r_{\rm in}=r_{\rm isco}$. 
When the sample spectra of observation A (the typical soft state) was fit by replacing {\sc SSsed} to 
 {\sc kerrbb} \citep{kerrbb}\footnote{{\sc kerrbb} was used under the condition of zero torque at the inner boundary, without limb-darkening and self-irradiation, and with a hardening factor of 1.7\citep{shimura1995}.} plus {\sc nthcomp}, 
the spin parameter was estimated to be $a^*=0.15\pm0.01$ with $\dot{m}=4.37\pm0.07$, which was
exactly same as that obtained by $r_{\rm in}$ in the {\sc SSsed} model.

We examined the data from the SIMS.
We allow both thermal and non-thermal Comptonisation, though the thermal Comptonisation is not always required. Figure~\ref{fig:spec_SSsed}b illustrates the SIMS spectrum (observation B), where the weaker and steeper ($\Gamma_{\rm th}=4$) Comptonisation (green) is seen between the outer disc (magenta) and the flatter non-thermal corona (blue). 
 As seen in this figure, the additional steep component is much weaker and steeper than the flatter Comptonising component, and thus it was not able to be distinguished whether it is thermal (with a cut-off) or non-thermal (without a cut-off). Instead, we fixed $kT_{\rm e}$ at 9 keV, which is an average value of the thermal Comptonisation component in the HIMS (see section \ref{sub:hims}).
 It is important to note that the uncertainty of $kT_{\rm e}$ does not affect the fitting result. 
  
In the SIMS, the values of $r_{\rm in}$ were found to be $\sim 4.5r_{\rm g}$ on average.
Details of $r_{\rm in}$ in the SIMS are presented in Appendix~\ref{app:rin_sims}.
The value of $4.5r_{\rm g}$ is slightly smaller than 
$r_{\rm in}$ of (5.0--$5.5)r_{\rm g}$ in the soft state, and the difference is statistically significant, since
$\chi^2_\nu$ values become much worse if spectra in the SIMS were fit with the {\sc SSsed} by fixing $r_{\rm in}$ at $5.5r_{\rm g}$. This could be a real effect, where the non-thermal region is given additional emissivity from magnetic connection across the ISCO, but it seems more likely that it arises from the model uncertainties and approximations
(e.g., slight difference from isotropic geometry in the corona region, geometry changes coupled with relativistic effects, or slight difference of inclination angle).
Consequently, we refer the average value of $4.5r_{\rm g}$ as an apparent
reference value for constant $r_{\rm in}$ (i.e. similar to $r_{\rm isco}$) in the SIMS.
This difference, which is  up to 20\%, could be the systematic uncertainty to estimate absolute values of 
corona radius $r_{\rm cor}$.


\subsection{The HIMS }
\label{sub:hims}
\begin{figure*}
\begin{center}
\includegraphics[width=17.5cm]{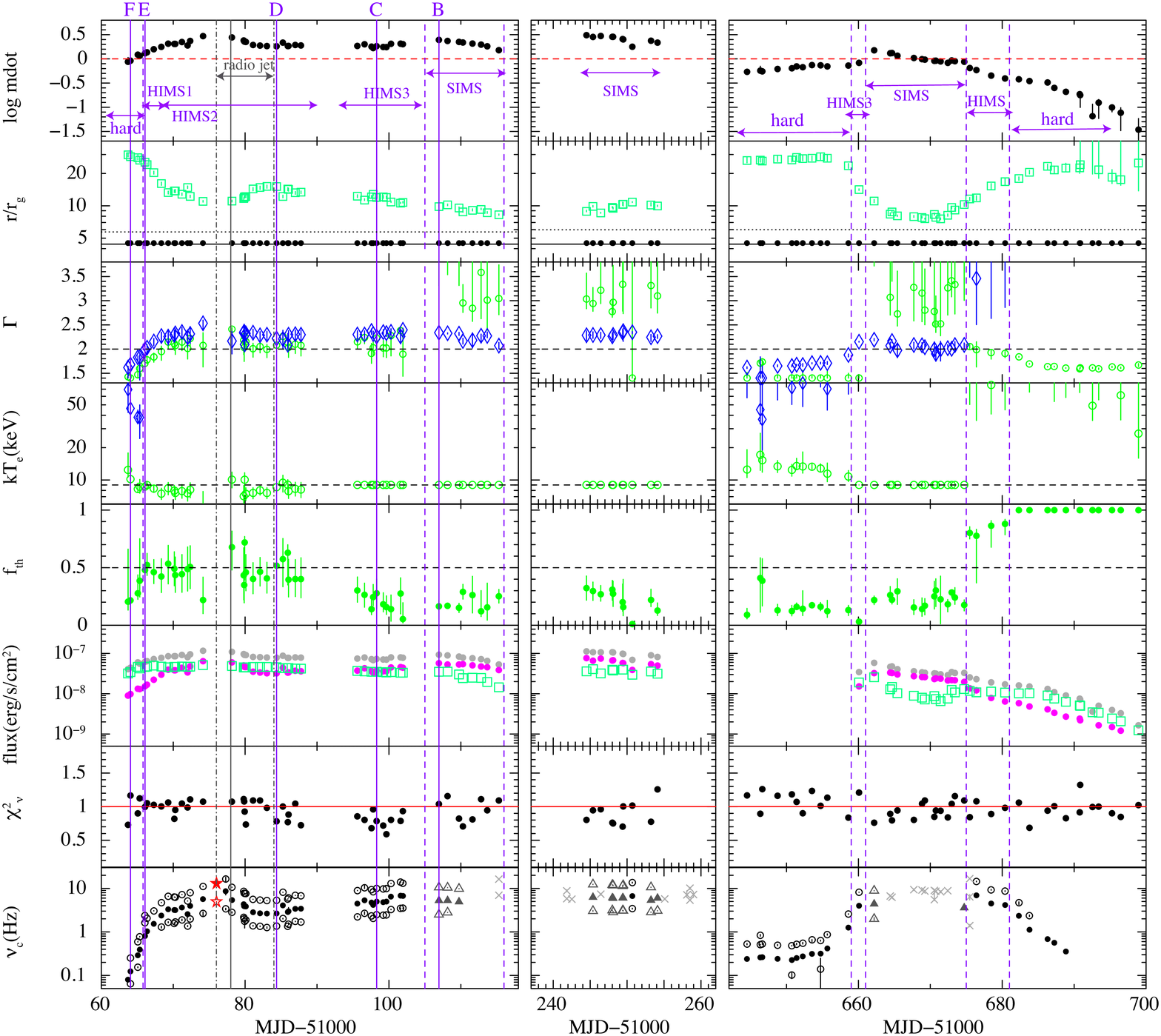}
\end{center}
\caption{Time evolution  of the best-fit parameters  of the {\sc SSsed}. 
Parameters shown are 
(1st panel) $\log \dot{m}$, with a horizontal dashed lines indicating $\dot{m}=1$. (2nd panel) The corona radii $r_{\rm cor}$ (open square) and  $r_{\rm in}$ (black dots, fixed at $4.5r_{\rm g}$). The $r_{\rm cor}$ values can be compared with 
the effective radius $r_{\rm max}(= (7/6)^2r_{\rm in}=1.36r_{\rm in})$ shown with the dotted line.
At $r_{\rm max}$, the local emissivity $\sigma T_{\rm eff}^4$ becomes maximum in the framework of the Shakura-Sunyaev standard disc.
(3rd panel) the photon index of the non-thermal ($\Gamma_{\rm nth}$, blue diamonds) and 
thermal ($\Gamma_{\rm th}$, green circles). (4th panel) Electron temperature of thermal Comptonising corona. 
(5th panel) The fraction of thermal Comptonising corona, $f_{\rm th}$. 
(6th panel) Bolometric flux values of the disc (filled magenta circles), Comptonising corona (open squares) and total (filled grey circles). 
(7th panel) Reduced chi-squared values $\chi^2_\nu$, with horizontal solid line indicates $\chi^2_\nu$ of 1,
(8th panel) QPO frequencies repeated from Fig.~\ref{fig:history_12}.
Vertical solid lines indicate the locations of observations B--F, and the vertical dashed lines distinguish 
the spectral states, and
vertical dash-dotted lines indicate the times at which the strong radio jet was detected, with maximum on MJD 51077--51078 \citep{hannikainen2009} as solid grey line. }
\label{fig:par_SSsed_hard}
\end{figure*}

The HIMS spectra required the inclusion of both thermal and non-thermal Comptonisation components. 
The examples of spectral fits of HIMS data are shown in Fig.~\ref{fig:spec_SSsed}c-\ref{fig:spec_SSsed}e.
Figure \ref{fig:par_SSsed_hard} presents the evolution of the spectral parameters in the HIMS and hard state.

The outer radius of the disc-corona region, $r_{\rm cor}$, was observed to be much larger than that in the soft state and SIMS.
As a result, 
and the temperature of outer thin disc, $T(r_{\rm cor})$, is 
too low to be constrained in the energy range of the PCA, despite the high luminosities.
Therefore, throughout the spectral fit,
the values of $r_{\rm in}$ were fixed to be an average value of $r_{\rm in}$ 
 in the SIMS, $4.5r_{\rm g}$ (see section~\ref{sub:soft-sims}).
Moreover, to achieve satisfactory $\chi^2_\nu$ values, 
the hot Comptonising component with thermal cut off (represented by the green lines in Fig.~\ref{fig:spec_SSsed}c--\ref{fig:spec_SSsed}e) 
is always necessary, and $f_{\rm th}$, $\Gamma_{\rm th}$, and $kT_{\rm e}$ were treated as a free parameter.
In this figure, the same SIMS data were analysed in the same manner as described in section~\ref{sub:soft-sims}, but with a fixed value of $r_{\rm in}$ at $4.5r_{\rm g}$.
This was done to facilitate a comparison of the results with those obtained in the HIMS and hard state, without considering uncertainties.

As seen in the second bottom panels of Fig.~\ref{fig:par_SSsed_hard}, 
almost all the data points in the HIMS were successfully reproduced with the {\sc SSsed}, 
except for 
the data obtained around a strong radio jet event on MJD 51076--51077\footnote{The fits were not improved even if the the value of $\Gamma_{\rm th}$ was not constrained. } \citep{hannikainen2009}. 
These data probably indicate
that there are additional processes
from the jet, which are not included in the baseline {\sc SSsed} model. We thus exclude these data from our discussion based on the {\sc SSsed} model.

During the first outburst in the HIMS phase, 
the size of the corona region, $r_{\rm cor}$, was
$\sim 25r_{\rm g}(=6r_{\rm in})$  at the onset (MJD 51066), decreasing to 
$\sim 16r_g(\sim 4r_{\rm in})$ on MJD 51068.
Except for the radio jet duration, it underwent a change to approximately (11--15)$r_{\rm g}$ by MJD 51087, gradually decreasing to (11--13)$r_{\rm g}$ during the softer HIMS phase (MJD 51095--51101), 
and eventually reaching $(8-10)r_{\rm g}$ in the SIMS phase (MJD 51106--51115).
From MJD 51066 to 51087, the contributions of thermal and non-thermal Comptonisation were almost equal, with $f_{\rm th}$ ranging from 0.4 to 0.6. However, from MJD 51095 to 51101, shortly before the start of the SIMS phase, the value of $f_{\rm th}$ decreased to 0.1--0.2.
Therefore, we categorise the epochs into HIMS1 (MJD 51066--51068), HIMS2 (MJD 51068--51087), 
and HIMS3 (MJD 51087--51095), ranging from hard to soft.
In the second outburst, the HIMS data on MJD 51660 are consistent with those of HIMS3 based on the values of $f_{\rm th}$ and $r_{\rm cor}$. Additionally, the data on MJD 51675--51680 are likely in the HIMS phase but are difficult to distinguish due to their faintness.

During the HIMS1 phase of the first outburst, the non-thermal Comptonising component was observed to have a spectral index of $\Gamma_{\rm nth}\sim 2.0$, which then evolved to become steeper, reaching values of $2.2\sim 2.3$. This steeper spectral index was maintained throughout the HIMS2 and HIMS3 phases.
Similarly, the thermal Comptonisation component showed variations, starting with a harder spectrum characterised by 
$\Gamma\sim 1.7 $ and $kT_{\rm e}\sim9$~keV, and then transitioning to a softer spectrum with $\Gamma=2$--2.2.
During the fitting of the HIMS3 data, the weaker thermal component did not allow for a reliable determination of $kT_{\rm e}$, so it was fixed at an average value of 9~keV.
During HIMS1, $y$-parameter and optical depth $\tau_{\rm es}$ of thermal Comptonising corona
varied from  $\tau_{\rm es}\sim 7$ and $y\sim 3$ to $\tau_{\rm es}\sim 5$ and $y\sim 1$. 

\subsection{The hard state : rising and decaying phases }
\label{sub:hard}
The hard state spectra in the bright 
rising phase of the outbursts (until MJD51065 in the first outburst and MJD51644--51655 in the second outburst) posed challenges when fitting them using the framework of thermal and non-thermal Comptonisation. The reduced chi-squared values were relatively high with hard spectral index of $\Gamma_{\rm nth}\sim1.7$, with $\chi^2_\nu$ value of 1.4 with d.o.f. of 114 on MJD51064 
(observation F) in the rising phase of the first outburst, and $\chi^2_\nu$ values ranging from 1.4 to 1.8 with d.o.f. of 108 on MJD51644--51655 in the second outburst.
There is no clear evidence of a disc in these spectra, so we again
fix $r_{\rm in}$ to $4.5r_{\rm g}$ as seen in the SIMS 
(see section~\ref{sub:soft-sims}). 

There was no clear evidence of a non-thermal power-law-like Comptonising component in these spectra. Instead, the dominant spectral component appeared to be hot thermal Comptonisation with a relatively hard photon index ($\Gamma<2$). However, these spectra could not be adequately reproduced using a single Comptonisation model (see details in Appendix~\ref{app:diskbb-nthc}).
Previous studies of the bright hard state by \cite{kawamura2022,kawamura2023} have suggested the presence of multi-temperature Comptonisation to explain similar complex spectral features, and \cite{yamada2013} also suggested the two temperature Comptonisation for Cyg X-1 in the hard state.

Therefore, we attempted to fit these spectra with the same {\sc SSsed} model but allowing the temperature ($kT_{\rm e}$) of both Comptonising components to vary freely. As a result, the spectra were successfully reproduced by employing two thermal Comptonisation components: one with a higher temperature 
($kT_{\rm e}=40$--70~keV) and a photon index of $\Gamma=1.6$--1.8 (i.e. $\tau_{\rm es}=3$--2, $y=3$--2), and the other with a lower temperature 
($kT_{\rm e}\sim10$~keV) and a photon index of $\Gamma=1.4$--1.6
(i.e. very large $\tau_{\rm es}=9$--7 and $y=7$--4). 
Figure~\ref{fig:spec_SSsed}f shows an extreme case of this spectral fit taken from observation F.
While the presence of two thermal Comptonisation components in the spectra of the bright hard state during the rising phase of the outbursts provides insight into the evolution of non-thermal Comptonisation, investigating the detailed structure of the hard state is beyond the scope of this paper. Instead, we utilised the estimated values of $r_{\rm cor}$ obtained from the improved spectral fits as a result of this analysis.

During the decaying phase of the outburst, the spectra in the low/hard state were successfully reproduced using a single hot thermal Comptonisation component. As the system transitioned from the HIMS to the hard state (between MJD 51680 and 51682), the spectral characteristics changed, with a dominant hot Comptonising component and no significant contribution from a steep non-thermal hard X-ray component originating from the disc-corona region. The spectra were well described by a single hot Comptonisation component with a photon index of $\Gamma=1.6$--1.7 and electron temperature of $kT_{\rm e}\ge 50$~keV. The coronal radii were found to extend further as $20r_{\rm g}$ to $30r_{\rm g}$. However, during this phase, the accretion rate ($\log\dot{m}$) was determined to be less than $-0.5$, corresponding to only 3\% of the Eddington luminosity. As a result, it is expected that the ADAF \citep{adaf} is realised instead of the assumed Shakura-Sunyaev emissivity.

\subsection{Summary of model parameters from the fits}

Figure \ref{fig:par_SSsed_hard} shows how the derived parameters of the {\sc SSsed} 
change across the outburst. There is a very nice smooth variation in $r_{\rm cor}$ with time
despite the large changes in spectral shape, with Comptonisation changing between thermal and non-thermal. 
The fits are generally very good (the very poor fits around a strong radio jet event on MJD 51076--510775 noted above 
are excluded from the analysis).

\section{Discussion}
\subsection{Anti-correlation between the corona size and LF-QPOs}

\begin{figure}
\begin{center}
\includegraphics[width=8cm]{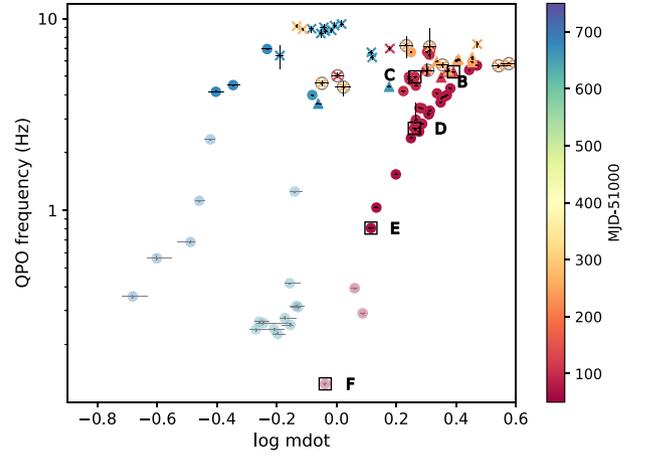}
\end{center}
\caption{QPO centroid frequency, $\nu_{\rm c}$, versus $\log \dot{m}$. Clearly there is the same 'parallel tracks' behaviour as in Fig.~\ref{fig:rate-fc}b of QPO centroid versus intensity, as $\dot{m}$ is closely related to intensity.  
The locations of observations B--F are indicated with open squares. 
Values of $r_{\rm in}$ in the hard state, HIMS, and SIMS are fixed at  $4.5r_{\rm g}$, 
 and the soft state data (without fixing $r_{\rm in}$) are identified with grey open circles.
 Type-C and Type-B QPOs are identified with circles and triangles, respectively, while Type-A QPOs and the other unclear structures are
identified with crosses. Colour-coded according to observation dates of ${\rm MJD}-51000$.}
\label{fig:mdot-qpo}
\end{figure}

We now explore how the QPO depends on the {\sc SSsed} model parameters. 
Figure~\ref{fig:mdot-qpo} shows the QPO frequency versus the overall mass accretion rate, $\log \dot{m}$. This shows the same parallel track behaviour as in Fig.~\ref{fig:rate-fc}b as the derived $\dot{m}$ is closely linked to the observed intensity. 
Instead, it is clear that the QPO frequency depends more tightly on the spectral shape (Fig.~\ref{fig:rate-fc}c), which is determined mainly by $r_{\rm cor}$ in the {\sc SSsed} model. 

\begin{figure*}
\begin{center}
\includegraphics[width=17cm]{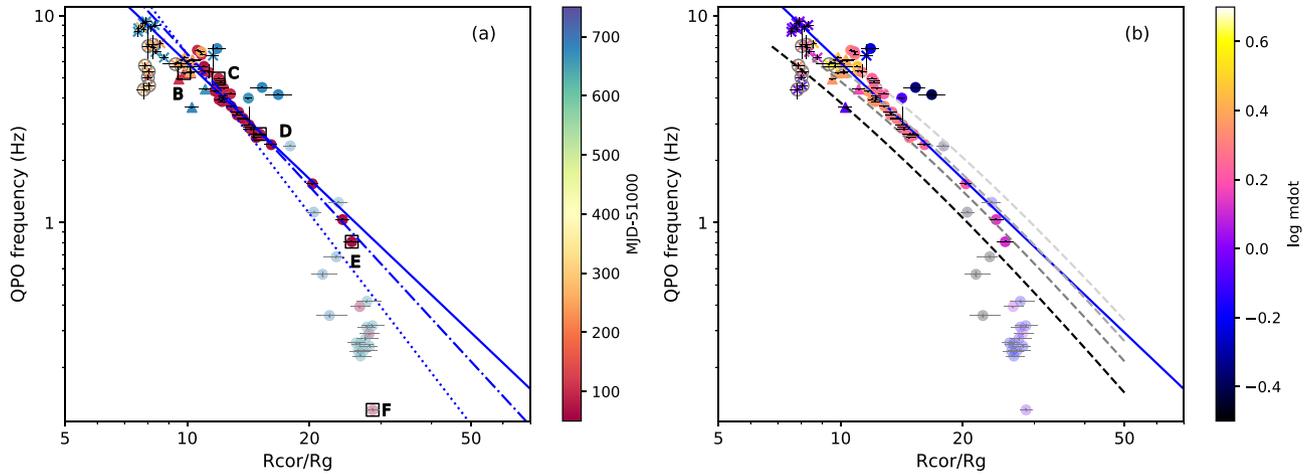}
\end{center}
\caption{QPO centroid frequency, $\nu_{\rm c}$, versus outer radius of Comptonising region, $r_{\rm cor}/r_{\rm g}$.
 Values of $r_{\rm in}$ in the hard state, HIMS, and SIMS are fixed at $4.5r_{\rm g}$, 
and the soft state data (without fixing $r_{\rm in}$) are identified with grey open circles.
The data points from the hard state shown in fainter colour. 
QPO types are identified with the same symbols in Fig.~\ref{fig:mdot-qpo}.
(a) Colour-coded according to observation dates of ${\rm MJD}-51000$.
Blue lines represent the best-fitting power-law.
The blue dotted line (all the data), solid line (without the hard state data), and dash-dotted line (the HIMS data alone) represent
$\nu_{\rm c}=2359\cdot r_{\rm cor}^{-2.559} $, $\nu_{\rm c}=434\cdot r_{\rm c}^{-1.865}$, and $\nu_{\rm c}=874\cdot r_{\rm cor}^{-2.126}$, 
respectively.
(b) Colour-coded according to $\log\dot{m}$.
Overlaid are the relations between  LT precession frequency and the outer radius of a hot flow (taken from Fig.~5 in 
IDF09) for spins of $a^\ast = 0.3$ (black),  $0.5$ (dark grey), $0.7$ (light grey) and 0.998 (lighter grey),
shifted up by a factor of $1.11(=10M_\odot/9M_\odot)$. }
\label{fig:r-fc}
\end{figure*}

Figures~\ref{fig:r-fc}a  presents the QPO centroid frequency plotted against the estimated $r_{\rm cor}$. This figure clearly illustrate the strong anti-correlation between $\nu_{\rm c}$ and $r_{\rm cor}$ throughout all phases of the outburst. In order to account for potential systematic uncertainties in determining $r_{\rm cor}$ during the hard state, the data points corresponding to the hard state are displayed with fainter colours.
It is important to note that the low/hard state with $\log\dot{m}<-0.5$ likely exhibits the influence of an ADAF, which is not accounted for in the {\sc SSsed} model. Similarly, the bright/hard state displays a complex corona structure (see section~\ref{sub:hard}). Additionally, in the hard state, the reverberation signal suggests a strong enhancement of outer disc emission due to reprocessing caused by irradiation from Comptonising emission. Such reprocessing effects are not included in the {\sc SSsed} model, which may lead to inaccuracies.

By excluding the data points from the hard state, we find that the anti-correlation between $r_{\rm cor}$ and $\nu_{\rm c}$ can be approximated by power-law relationship: $\nu_{\rm c}=434\cdot (r_{\rm cor}/r_{\rm g})^{-1.865}$~Hz. 
Similarly, the relation ship of $\nu_{\rm c}=874\cdot (r_{\rm cor}/r_{\rm g})^{-2.126}$ was obtained by the HIMS data alone\footnote{
Power-laws of $\nu_{\rm c}=2359\cdot (r_{\rm cor}/r_{\rm g})^{-2.559}$~Hz  was obtained with all the data including the hard state.}.
Indeed, the observed anti-correlation between $\nu_{\rm c}$ and $r_{\rm cor}$ provides an empirical description of the relationship between QPO centroid frequency and the extent of the hard X-ray corona. This relationship holds consistently across different outbursts, as represented by the variation in colour, as well as different types of QPOs, as represented by the different symbols used in the plot. This suggests a robust connection between the QPO properties and the characteristics of the corona, emphasising the significance of the corona in driving the QPO behaviour.

This conclusion is consistent with a previous study by
\cite{marcel2020} which also found that the frequency of QPOs is related to the energy-conservation of  an inner jet-emitting disc \citep{ferreira2006,marcel2019} and an outer 
standard disc, based on radio, X-ray and $\gamma$-ray observations.

\subsection{Comparison between the observation and the LT prediction}

In order to compare the observed anti-correlation between $\nu_{\rm c}$ and $r_{\rm cor}$ with theoretical predictions proposed to explain the low-frequency quasi-periodic oscillations (LF-QPOs), we focus on investigating the relativistic Lense-Thirring (LT) precession scenario in this paper.

The relativistic LT precession was first proposed as a candidate to cause LF-QPOs by \citet{stella1998}, 
and it has been extensively studied since by many authors including IDF09 and \citet{motta2014_1655}.
IDF09 coupled the mechanism with the truncated disc/hot inner flow geometry to predict an
anti-correlation between $\nu_{\rm c}$ and the inner radius of the truncated disc (which corresponds to the outer radius of geometrically thick hot inner flow).
Figure~\ref{fig:r-fc}b shows the same $\nu_{\rm c}$-$r_{\rm cor}$ plot as in Fig.~\ref{fig:r-fc}a but
colour coded by $\log\dot{m}$.

In Fig.~\ref{fig:r-fc}b, the predicted $\nu_{\rm c}$-$r_{\rm cor}$ relations based on LT precession scenario by  IDF09 are overlaid with dashed lines for spin parameters of $a^*=0.3$, 0.5, 0.7 and 0.998 (lower to upper lines).
The predicted lines are taken from Fig.~5 in IDF09
(though we note that these are for a fiducial mass of $10M_\odot$ rather than the $9M_\odot$ used here). 
The observed $\nu_{\rm c}$--$r_{\rm cor}$ relations are consistent with the 
LT prediction.
This result is strongly supporting LT precession as a probable candidate 
for the origin of the LF-QPOs.
However, there is still some scatter in the data points around the predicted lines. 
They do not exactly pick out a single spin parameter track in the LT models, though they are supposed
to concentrate on the lines $a^\ast=0.5$--0.7.

This is not surprising given the LT model simplicity in IDF09, where the precessing Comptonised region is assumed to have fixed scale height for all $\dot{m}$. It seems more likely that the Comptonisation geometry is changing during the state transition, so that the scale height decreases as the Compton cooling becomes larger.
This changes the predicted LT precession frequency, as well as changing the spectra (by changing the inclination dependence and relativistic corrections)  in a way which is not included in {\sc SSsed}. 
Moreover, the {\sc SSsed} model also does not include
the jet power, yet there is a steady compact jet in the bright hard state and HIMS spectra which is powered by the accretion flow (e.g. \citealt{fender2005}), and there are discrete blobs ejected at the HIMS/SIMS transition (e.g. \citealt{wood2021}). 

Perhaps a more serious issue is that the disc-dominated spectra in this source give a black hole spin of $a^*\sim 0.15$ when fit with {\sc kerrbb} or $r_{\rm in}=5.5r_{\rm g}$ for the 
assumed binary parameters of $d=4.4$~kpc, $i=75^\circ$ and $M=9M_\odot$ (see section~\ref{sub:soft-sims}).
This seems very low to cause the strong precession required to produce the large QPO amplitude observed.
However, we note that there are uncertainties on these values, and closer distance and higher mass results in larger black hole spin.
Including these system uncertainties gives a wider range of 
spin estimates from the disc-dominated spectra, of $0.1< a^\ast< 0.7$ \citep{davis2006}
or $-0.11< a^\ast<0.76$ \citep{steiner2011} with \citet{connors2020} using the spin of 
$\sim 0.5$ to encompass all current estimates. 

Finally, it is important to note that
while the role of LT precession has been primarily proposed to explain Type-C QPOs, there is no significant observational difference in the behavior of Type-B (and A) QPOs in the $\nu_{\rm c}$--$r_{\rm cor}$ relation
in view of the frequencies. Our analysis indicates that both types of QPOs are consistent with the LT precession scenario, with Type-B (and A) QPOs appearing at the high $\nu_{\rm c}$ and small $r_{\rm cor}$ end of the $\nu_{\rm c}$--$r_{\rm cor}$ relation, although there may be some scatterings.

\subsection{Issues for LT precession models}
To construct the {\sc SSsed} model, we utilised the passive disc scenario to provide seed photons to the Comptonising component. It should be noted that LT precession, being a vertical precession, so is challenging to occur if there are obstacles present in the mid-plane.
However, in the calculations of the {\sc SSsed} model, the passive disc primarily serves as a source of seed photons for the coronae. The specific details of the seed photon structure have minimal impact on the overall results. Hence, slight variations or changes in the seed photon structure do not significantly influence the obtained results in the model. 
The presence of strong non-thermal Comptonisation in the HIMS and SIMS indicates the need for a substantial source of seed photons, 
which can be from strong external irradiation from the 
truncated disc and/or the steep Comptonisation can provide seed photons for the thermal Comptonisation. Alternatively, the 
passive disc can have a clumpy structure, so that the hot plasma can precess through the mid-plane holes. 
Such a clumpy structure may be consistent with optically thick blobs predicted in GRMHD simulations \citep{liska2022}. 

In fact, there was no significant difference in estimating the size of the outer corona, $r_{\rm cor}$, 
when we changed slightly the corona geometry. For example, we tried to fit the data with truncated disc-(non-thermal) corona and inner hot flow geometry, the obtained $\nu_{\rm c}$-$r_{\rm cor}$ relation are almost the same and still consistent with the LT prediction. 
Therefore, we can conclude that the estimation of $r_{\rm cor}$ is quite robust, as it is primarily determined by the combination of the outer disc's shape and luminosity, as well as the relative luminosity of the outer disc compared to the Comptonisation components.

In addition,
LT precession of the entire Comptonised flow matches (to zeroth order) all the QPO types seen here, not just the  Type-C QPOs, with
Type-B and Type-A QPOs appearing at 
the smallest $r_{\rm cor}$ values in these data rather than being offset as might be the case if these were really from a different mechanism/geometry (see e.g. the review by \citealt{ingram2019}).

Totally, 
the observed $\nu_{\rm c}$--${\rm r_{\rm cor}}$ relation is consistent with its theoretical 
prediction calculated by IDF09.
However, it is important to note that the scenario of LT precession as the origin of LF-QPOs is 
still being questioned by some authors.
For example, \citet{marcel2021} demonstrated that the accretion flow can reach sonic or supersonic speeds in the bright hard state, which prevents the 'wave-like' regime required for solid-body precession to cause LF-QPOs. This can also result in strong changes in the predicted QPO frequency due to the magnetic interaction of the precessing flow with the outer disc \citep{bollimpalli2023}.

Therefore, further theoretical studies including GRMHD simulations are necessary to understand the LF-QPO behaviour. 
Nevertheless, the observed $\nu_c$--$r_{\rm cor}$ relation can provide a useful guideline to investigate the validity of theoretical models for LF-QPOs.

\section{Summary and Conclusion}

We developed the spectral model {\sc SSsed} by modifying the multicomponent  AGN model {\sc agnsed} to tailor it for  stellar black hole binaries. We  applied it to the
X-ray spectra of XTE~J$1550-564$ observed with RXTE to determine the size of Comptonising corona in different states. 
Except for the data obtained around a strong radio jet event on MJD 51076--51077 and
during the rising phase of outburst, 
the {\sc SSsed} model successfully reproduced the spectra of XTE~J$1550-564$.

In the HIMS,  both thermal ($kT_{\rm e}=8$--20~keV) and non-thermal ($\Gamma=2$--2.4 without clear cut-off energy) corona were required to reproduce 
the Comptonising component.
The outer corona radii were found to vary from $10r_{\rm g}$ to $25r_{\rm g}$. 
In the SIMS, though the hard emission was dominated by the non-thermal Comptonising component, 
an additional steeper ($\Gamma\le2.5$) Comptonising components improved the spectral fits with fraction of $f_{\rm th}=0.1$--0.3.
The corona radii were found at (8--$10)r_{\rm g}$ .

The hard state spectra did not require the non-thermal Comptonising component.
On one hand, in the hard state during the decaying phase of the 2nd outburst, 
the spectra were well represented by a single hot Comptonising component with $\Gamma=1.6$--1.8 and $kT_{\rm e}\sim 50$~keV at $r_{\rm cor}>20r_{\rm g}$. 
On the other hand, spectra in the bright hard state during the rising phase of the outbursts were difficult to be fit 
within the framework of thermal and non-thermal Comptonising corona.
Instead, the spectra were reproduced with a two-temperature thermal Comptonizing corona, which is consistent with the findings suggested  by \cite{kawamura2022,kawamura2023}. 

By comparing the obtained corona outer radii $r_{\rm cor}$ to the centroid LF-QPO frequency, $\nu_{\rm c}$, 
we found a tight anti-correlation between $\nu_{\rm c}$ and $r_{\rm cor}$, 
described approximately as 
$\nu_{\rm c}=434\cdot r_{\rm c}^{-1.87}$ (without the hard state data) or $874\cdot r_{\rm cor}^{-2.13}$ (with the HIMS data alone).
This relation is 
in remarkably good qualitative and quantitative agreement with the prediction of LT precession (IDF09). 
We conclude that the intermediate states, which show both strong disc emission and strong LF-QPOs, provide the best data 
to test QPO formation models despite the complexity of these spectra.

\section*{Acknowledgements}
AK thanks S. Mineshige and T. Kawashima for their helpful discussions.
AK acknowledges the support received from the discretionary fund of the President of Shibaura Institute of Technology.
CD acknowledges
the Science and Technology Facilities Council (STFC) through grant
ST/T000244/1 for support, and University of Tokyo Kavli IPMU.
We also thank R. Connors for his detailed reading as a reviewer.

\section*{Data Availability}
The RXTE data underlying this article are available in NASA’s HEASARC archive site, at https://heasarc.gsfc.nasa.gov/docs/archive.html

\bibliographystyle{mnras}

\appendix 

\section{Spectral parameters of the {\sc SSsed} model}
\label{app:SSsed_par}

Spectral parameters of the {\sc SSsed} model are summarised in Table~\ref{tab:parameter}.
The model has some switching parameters. 
If parameter 6 is negative, the model gives the inner hot Comptonisation component.
If parameter 7 is negative, the model gives the Comptonisation component in the passive-disc corona region. If parameter 9 is negative, the model gives the outer disc. If parameter 12 is $-1$, the code will use the self gravity radius as calculated from \citet{laor1989}.
Colour correction is included when parameter 13 is fixed to be 1, while it is not included when this parameter is fixed to be 0. 
For BHB spectra, this parameter should be fixed to be 1.

\begin{table}
  \caption{Spectral parameters of the {\sc SSsed} model}%
  \begin{tabular}{lp{6.5cm}}
      \hline
      &  description\\
      \hline
      par1 & Mass, black hole mass in solar masses\\ 
      par2 & Dist, comoving (proper) distance in kpc \\
      par3 & $\log\dot{m}$, $\dot{m}=\dot{M}/\dot{M}_{\rm Edd}$ where $\dot{M}_{\rm Edd}c^2=L_{\rm Edd}$ \\
      par4 & $r_{\rm in}$, inner most radius of the accretion flow in $r_{\rm g}$\\
par5 & $\cos i$, inclination angle of the disc\\      
par6&$kT_{\rm e,th}$, electron temperature for thermal corona in keV. If this parameter is negative, the model gives the inner hot Comptonisation component.  \\ 
par7&$kT_{\rm e,nth}$, apparent electron temperature for non-thermal corona in keV which is recommended to be fixed at 300~keV to mimic non-thermal electron distribution. If this parameter is negative, the model gives the Comptonisation component in the passive-disc corona region.
\\ 
par8&$\Gamma_{\rm th}$, photon index of inner hot corona.
 If this parameter is negative then only the
inner Compton component is used.\\
par9&$\Gamma_{\rm nth}$, photon index of disc-corona. If this parameter is negative, the model gives the outer disc. \\
par10& $f_{\rm th}$, fraction of the hot Comptonising component to the total Comptonisation \\
par11& $r_{\rm cor}$, outer radius of the disc-corona region in $r_{\rm g}$\\
par12& $\log r_{\rm out}$,  outer radius of accretion disc in $r_{\rm g}$. If this parameter is $-1$, the code will use the self gravity radius as calculated from \citet{laor1989}\\
par13&redshift, must be fixed\\
par14&$f_{\rm col}$, switching parameter for colour correction (0: no colour correction, 1: colour correction factor is calculated by the same way as {\sc optxagnf}  \citep{optxagnf}\\
par15&norm, must be fixed at 1\\
       \hline
    \end{tabular}
    \label{tab:parameter}
\end{table}

\section{Second Compton component and reflection hump based on canonical {\sc diskbb} and {\sc nthcomp} modelling}
\label{app:diskbb-nthc} 

\begin{table*}
   \caption{The best-fitting parameters of `{\sc tbabs*(diskbb+kdblur*xilconv*nthcomp+Gaussian}'  model for observations A--F. The system parameters are fixed at the same values as Table~\ref{tab:spec_SSsed}. Errors are not estimated for observations D to F due to bad chi-squared values. $^\dagger$The best-fitting electron temperature of {\sc nthcomp} were found at their upper limit of 300~keV, and thus they were fixed. 
   $^*$Parameters hit the lower or upper limit. $^\#$Chi-squared values for the same fits without Gaussian component, and $F$-values calculated as 
   $F=(\Delta \chi^2/\chi^2_\nu)/\Delta \nu$, representing the improvement due to the Gaussian component.}
  \begin{tabular}{p{2cm}p{2cm}p{2cm}p{2cm}p{2cm}p{2cm}p{2cm}}
      \hline 
Obs.             &A&B & C& D & E &F\\
state        &soft&  SIMS & HIMS3& HIMS2 & HIMS1 & hard\\
\hline
\multicolumn{7}{l}{parameters of {\sc diskbb}}\\
$kT_{\rm in}$(keV) & $0.828^{+0.002}_{-0.003}$&$1.008^{+0.007}_{-0.009}$& $0.83^{+0.02}_{-0.03}$&0.553& 0.844&0.941\\
$r_{\rm in}(r_{\rm g})$ &$50.7^{+0.6}_{-0.5}$ &$40.9^{+0.5}_{-0.8}$&$43^{+3}_{-2}$& 55.5& 2.95 & 6.5 \\
\multicolumn{7}{l}{parameters of {\sc nthcomp}}\\
$\Gamma$ & $2.05\pm0.02$ & $2.428^{+0.008}_{-0.014}$&$2.44^{+0.03}_{-0.02}$ & 2.33 &2.01 &  1.74\\
$kT_{\rm e}$(keV) & (300)$^\dagger$ & (300)$^\dagger$  & (300)$^\dagger$  & 29.9 &  21.6 & 38.8\\
norm& $0.138^{+0.010}_{-0.015}$& $2.18^{+0.07}_{-0.08}$ & $3.6^{+0.3}_{-0.2}$& 7.46&  2.47& 0.779\\
\multicolumn{5}{l}{parameters of {\sc kdblur}, {\sc xilconv} and Gaussian}\\
$r_{\rm ref,in}(r_{\rm g})$ &$2.2^{+0.6}_{-0.5}$ &$2.2^{+0.5}_{-0.4}$ & $2.4^{+0.6}_{-0.7}$ & 25.2 & 11.1& 8.15\\
rel\_refl & $1.8\pm0.2$ &$0.87^{+0.10}_{-0.14}$ & $0.94^{+0.16}_{-0.17}$ &  1.01  & 1.22 & 1.23\\
$\log\xi$& $3.01^{+0.06}_{-0.04}$ &$3.19^{+0.12}_{-0.07}$ & $1.73^{+0.11}_{-0.27}$ & 2.77 & 2.06 & 1.99\\
$E_{\rm c}$(keV)&$6.58^{+0.11}_{-0.13}$ & $6.59^{+0.10}_{-0.11}$ & $6.26^{+0.11}_{-0.06}$ & 6.20* & 6.20* &6.20\\
$\sigma$(keV)&$0.80^{+0.00*}_{-0.04}$ & $0.80^{+0.00*}_{-0.07}$ & $0.66^{+0.07}_{-0.12}$ &0.10* & 0.10* & 0.10*\\
norm ($\times 10^{-3}$)& $7.2^{+3.5}_{-0.5}$   & $38^{+2}_{-6}$ & $41\pm8$ & 3.45 & 7.48& 4.5\\
\hline
EW(eV)&  $200^{+97}_{-14}$ &$157\pm12$ &$190\pm37$ & 12 &37 & 53\\
\hline
$\chi^2$(dof) & 137.6(115)& 161.7(115) & 105.0(115) &222.9(114) & 445.4(114) & 209.4(114)\\
\hline \hline
$\chi^2$(dof)$^\#$  & 319.6(118)& 293.3(118) & 519.8(118) & 226.5(117) & 450.7(117) & 246.6(117)\\
$F$-value$^\#$($\Delta$dof)    & 50.7(3) & 31.2(3) & 151.4(3) & -- & -- & --\\
\hline
\end{tabular} 
\label{tab:spec_diskbb-nthc}
\end{table*}

\begin{table}
    \caption{Same as Table~\ref{tab:spec_diskbb-nthc} but with two Comptonising components. Spectral fits for observations D--F were shown. Symbols are the same as Table~\ref{tab:spec_diskbb-nthc}.}

  \begin{tabular}{p{2cm}p{1.6cm}p{1.6cm}p{1.6cm}}
      \hline 
Obs.            & D & E &F\\
state       & HIMS2 & HIMS1 & hard\\
\hline
\multicolumn{4}{l}{parameters of {\sc diskbb}}\\
$kT_{\rm in}$(keV) & $0.68^{+0.03}_{-0.04}$&$0.78\pm0.08$&$0.81^{+0.16}_{-0.21}$\\ 
$r_{\rm in}(r_{\rm g})$ &$46^{+10}_{-4}$& $<6.8$ &$8.1^{+8.3}_{-1.2}$\\
\multicolumn{4}{l}{parameters of {\sc nthcomp}}\\
$\Gamma$ & $2.00^{+0.06}_{-0.00*}$& $2.00^{+0.03}_{-0.00*}$&$1.710^{+0.018}_{-0.016}$\\
$kT_{\rm e}$(keV) &(300)$^\dagger$ & (300)$^\dagger$ &$63^{+28}_{-13}$\\
norm&$0.77^{+2.60}_{-0.06}$ & $1.29^{+0.17}_{-0.10}$ &$0.67^{+0.04}_{-0.08}$ \\
$\Gamma$ &$2.242^{+0.018}_{-0.014}$&$1.750^{+0.015}_{-0.029}$ &$1.4^{+0.04}_{-0.00*}$\\
$kT_{\rm e}$(keV) &$11.4^{+0.9}_{-1.5}$ & $9.22^{+0.17}_{-0.34} $  & $11.2\pm0.7$\\
norm& $4.5^{+0.4}_{-1.5}$ & $1.06^{+0.05}_{-0.09}$ &$0.099^{+0.049}_{-0.013}$\\
\multicolumn{4}{l}{parameters of {\sc kdblur}, {\sc xilconv} and Gaussian}\\
$r_{\rm ref,in}(r_{\rm g})$ &$30^{+21}_{-11}$ & $25^{+36}_{-9}$ &$13^{+8}_{-5}$  \\
rel\_refl &$0.67\pm0.06$ &$0.51\pm0.06$&$0.53^{+0.06}_{-0.07}$\\
$E_{\rm c}$(keV)&$6.20^{+0.05}_{-0.00*}$ & $6.31\pm0.09$ & $6.26^{+0.09}_{-0.00*}$\\
$\sigma$(keV)&(0.3)$^\dagger$ & (0.3)$^\dagger$ & (0.3)$^\dagger$ \\
norm ($\times 10^{-3}$)&$17\pm6$ &$14^{+4}_{-2}$ &$6.2^{+1.5}_{-1.7}$\\
\hline
EW(eV)&  $59\pm21$&$75^{+21}_{-11}$&  $74^{+18}_{-20}$\\
\hline
$\chi^2$(dof) &83.2(114)&106.6(114) & 134.4(113)\\
\hline \hline
$\chi^2$(dof)$^\#$ &104.1(116)&145.0(116) & 160.4(115)\\
$F$-value$^\#$($\Delta$dof)  & 14.3(2) & 20.5(2) & 10.9(2)\\
\hline
  \end{tabular} 
 \label{tab:spec_diskbb-nthc-nthc}
\end{table}

\begin{figure*}
 \begin{center}
\includegraphics[width=17cm]{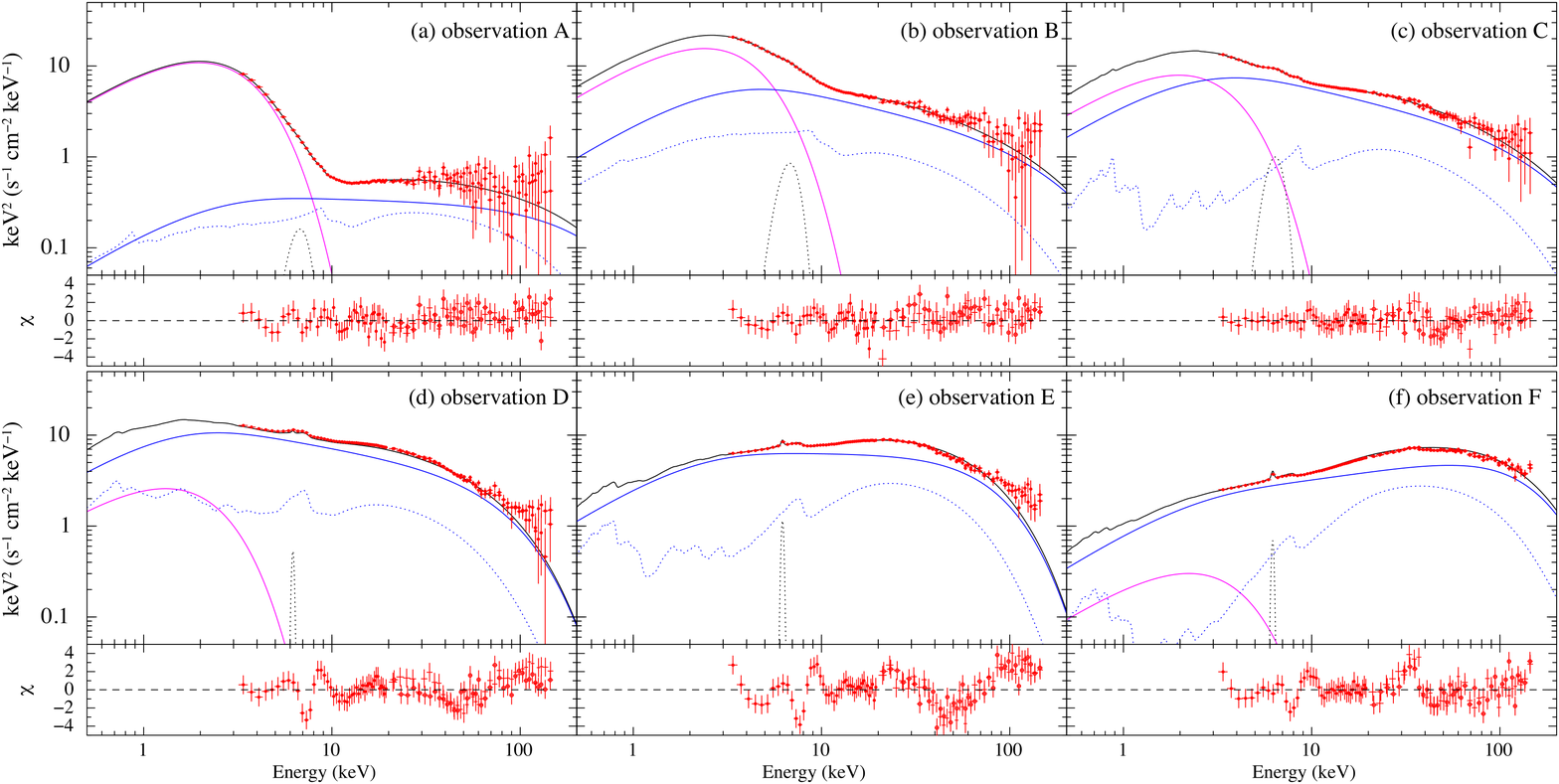}
\end{center}
\caption{Unabsorbed, deconvolved spectra of observations A--F fitted with `{\sc tbabs*(diskbb+kdblur*xilconv*nthcomp+Gaussian)'}.
{\sc diskbb} and {\sc nthcomp} components are coloured with magenta and blue, respectively. The reflected emission of {\sc nthcomp} is shown with 
blue dotted line. The model components are based on the parameters 
shown in Table~\ref{tab:spec_diskbb-nthc}.}
\label{fig:spec_diskbb-nthc-ref}
\end{figure*}

\begin{figure*}
 \begin{center}
\includegraphics[width=17cm]{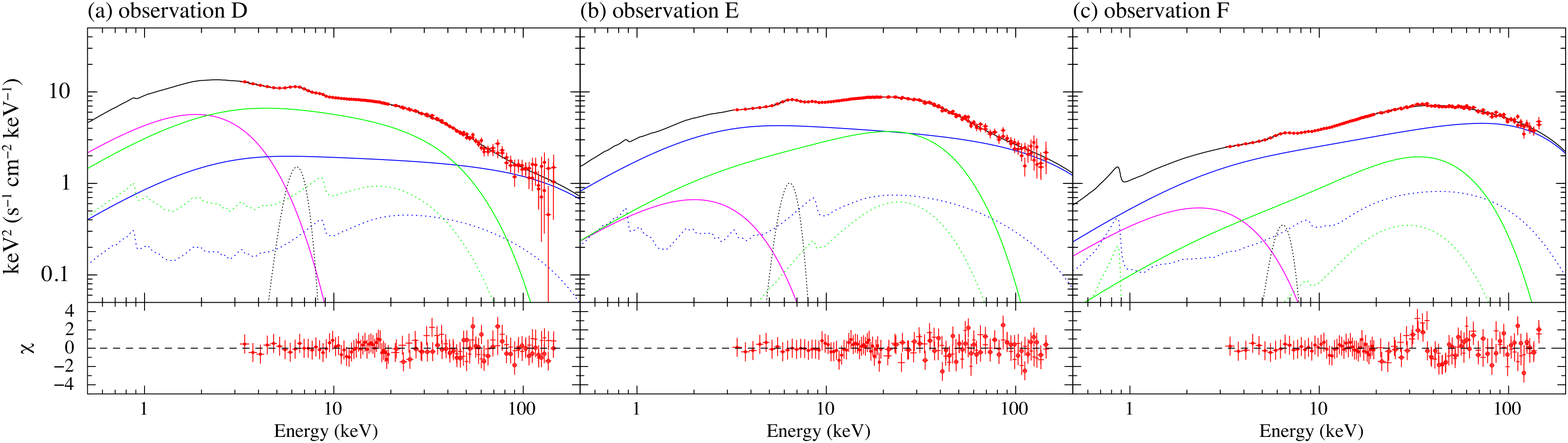}
\end{center}
\caption{Unabsorbed, deconvolved spectra of observations D--F fitted with  `{\sc tbabs*(diskbb+kdblur*xilconv*(nthcomp+nthcomp)+Gaussian)}'. 
{\sc diskbb}, higher and lower temperature {\sc nthcomp} components are coloured with magenta, blue, and green, respectively. 
The reflected emissions of two {\sc nthcomp} are shown with blue and green dotted lines. The model components are based on the best-fitting parameters 
shown in Table~\ref{tab:spec_diskbb-nthc-nthc}.}
\label{fig:spec_diskbb-nthc-nthc-ref}
\end{figure*}

We illustrate how the spectral fit changes with a single Comptonising component and two Comptonising components, including full reflection structures. The analysis is based on canonical {\sc diskbb} and {\sc nthcomp} modelling. 
The reflected emission of the Comptonising emission components is represented by using the {\sc xspec} convolution 
model {\sc xilconv} blurred by {\sc kdblur}. 
The convolution model {\sc xilconv}\footnote{https://heasarc.gsfc.nasa.gov/xanadu/xspec/models/xilconv.html}
is the update version of {\sc rfxconv}\citep{mari2011, done2006b} by C. Done, 
and describes an angle-dependent reflection from an ionized disc and includes parameters such as 
the relative reflection normalisation, {\sc rel\_refl}, ionization parameter, $\log \xi$, inclination angle, $\cos i$,
iron abundance, and cutoff energy $E_{\rm cut}$.  
The model combines an ionized disc table from the 
{\sc xilver} model \citep{garcia2013} with Compton reflection code by \cite{magdziarz1995}. 
The {\sc kdblur} model describes general relativistic blurring around a non-spinning black hole based on {\sc diskline} model \citep{fabian1989},

The description of {\sc xspec} is
\[tbabs*(diskbb+kdblur*xilconv*nthcomp+Gaussian)\]
where the seed photon temperature of {\sc nthcomp} was tied to $kT_{\rm in}$ of {\sc diskbb}, and the seed photon distribution type of {\sc nthcomp} 
was set to {\sc diskbb}. 
The parameters $E_{\rm c}$ and $\sigma$ of Gaussian are constrained to the range of 6.2--6.9~keV and 0.1--0.8~keV, respectively. 
The parameters for inclination, iron abundance, $E_{\rm cut}$, and outer radius of reflection for {\sc kdblur} and {\sc xilconv} were fixed at $75^\circ$, 1,  300~keV and $400r_{\rm g}$, respectively. The other parameters including
$\xi$-parameter, {\sc rel\_refl} and inner radius of reflection, $r_{\rm ref, in}$, were left as free parameters. 

Figures~\ref{fig:spec_diskbb-nthc-ref}a--\ref{fig:spec_diskbb-nthc-ref}f and Table~\ref{tab:spec_diskbb-nthc} show the resulting spectral fits for observations A--F.
For observations A--C, characterised by weaker {\sc nthcomp} components compared to observations D--F, 
satisfactory fits were achieved with $\chi^2_\nu$ in the range of 0.91--1.4.
While {\sc xilconv} reproduces not only the reflection continuum but also emission lines,
the iron line produced by the reflected emission is significantly blurred.
The observed iron line cannot be reproduced solely by the blurred reflection, and the additional Gaussians are statistically important. 
The significance of the Gaussians is examined by fitting with and without them. 
The chi-squared values without the Gaussian 
and $F$-values for improvement the fit with Gaussian were shown together in Table~\ref{tab:spec_diskbb-nthc}, where
$F$ is defined as $(\Delta\chi^2/\chi^2_\nu)/\Delta \nu$. 
It is possible that a large fraction of the iron line does not originate from the 
single reflection but rather from other sources such as winds and/or the other reflected emission from very outer disc as suggested by \cite{connors2019}. 

For observations D, E and F, the fit were considerably less satisfactory with $\chi^2_\nu$ of 2.0, 3.9, and 1.9, respectively.
Therefore, we proceeded to fit these spectra with '{\sc diskbb} plus two {\sc nthcomp}' incorporating full reflection.
For observations D and E, the spectra were well reproduced by introducing a second Compton component with
$\Gamma\ge 2.0$ and $kT_{\rm e}=300$~keV, indicating non-thermal Comptonisation.
The best-fitting parameters are presented in Table~\ref{tab:spec_diskbb-nthc-nthc}, and the corresponding spectra are presented in 
Figs.~\ref{fig:spec_diskbb-nthc-nthc-ref}a and \ref{fig:spec_diskbb-nthc-nthc-ref}b.
However, observation F (the bright hard state) 
exhibited a remarkably harder spectrum with an evident thermal cutoff, rendering the additional Compton component with
$\Gamma>2.0$ with $kT_{\rm e}=300$~keV ineffective in reproducing the spectra.
Consequently, when we employed two thermal Comptonised components for the fit, satisfactory results were achieved, as 
shown in 
Table~\ref{tab:spec_diskbb-nthc-nthc} and Fig.~\ref{fig:spec_diskbb-nthc-nthc-ref}c.

\section{The {\sc SSsed} fits with blurred reflection} 
\label{app:SSsed-ref}

\begin{figure*}
 \begin{center}
\includegraphics[width=17cm]{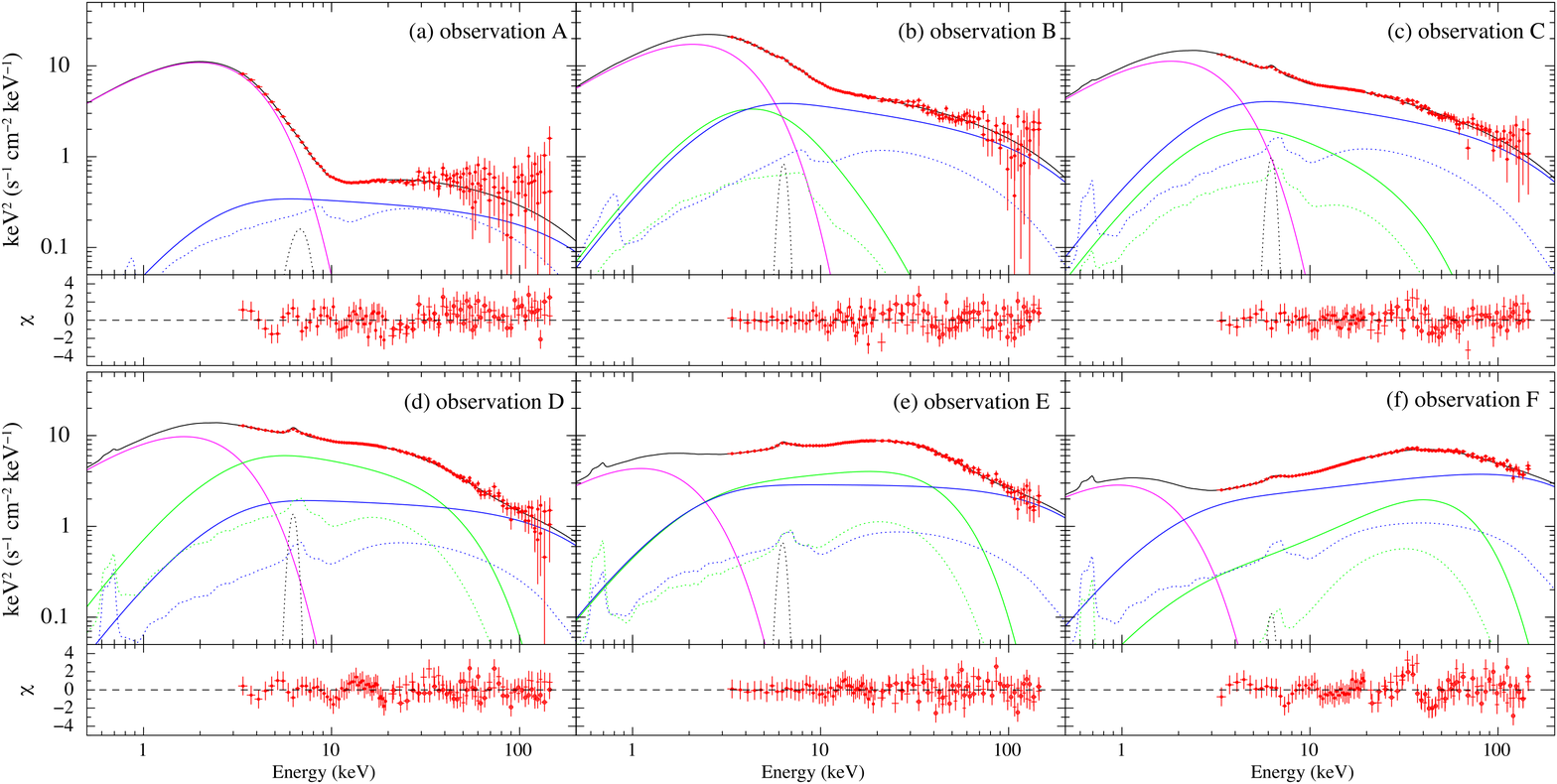}
\end{center}
\caption{Unabsorbed, deconvolved spectra of observations A--F fitted with {\sc SSsed} with blurred reflection by using {\sc kdblur} and {\sc xilconv}.
The model components are based on the best-fitting values shown in Table~\ref{tab:spec_SSsed-ref}.
Colours are the same as Fig.~\ref{fig:spec_SSsed}, and reflection components are shown with dotted lines. }
\label{fig:spec_SSsed-ref}
\end{figure*}

\begin{table*}
    \caption{The best-fitting parameters of '{\sc tbabs*(SSsed+kdblur*xilconv*SSsed$_{\rm coronae}$+Gaussian)}'.
   $^\ast$ The fitting values reach their lower or upper limits.
    $^\dagger$ The values in the parentheses are fixed. $^\S$The inner radius for reflection is tied to $r_{\rm cor}$.
    $^\ddagger$The values of iron abundance, outer radius for reflection and inclination angle are fixed at nominal values: i.e. 
  $Z_{\rm Fe}=1$, $r_{\rm ref,out}=400r_{\rm g}$, $i_{\rm xilconv}=75^\circ$.  $^\#$Same as Tables~\ref{tab:spec_diskbb-nthc} and \ref{tab:spec_diskbb-nthc-nthc}, Chi-squared values for the same fits without Gaussian component, and $F$-values calculated as 
   $F=(\Delta \chi^2/\chi^2_\nu)/\Delta \nu$, representing the improvement due to the Gaussian component.}
  \begin{tabular}{p{2cm}p{2cm}p{2cm}p{2cm}p{2cm}p{2cm}p{2cm}}
      \hline
Obs.             &A&B & C& D & E &F\\
state        &soft&  SIMS &HIMS3& HIMS2 & HIMS1 & hard\\
\hline
\multicolumn{7}{l}{parameters of the {\sc SSsed}}\\
$\log\dot{m}$   &$0.137^{+0.001}_{-0.013}$  &$0.36\pm0.03$&$0.206\pm0.003$&$0.204\pm0.006$& $0.037\pm0.009$&$-0.097^{+0.013}_{-0.014}$\\
  $r_{\rm in}(r_{\rm g})$ &$5.37^{+0.04}_{-0.02}$ &$4.64^{+0.24}_{-0.29}$&(4.5)$^\dagger$&(4.5)$^\dagger$& (4.5)$^\dagger$ & (4.5)$^\dagger$\\
 $r_{\rm cor}(r_{\rm g})$ & $6.702^{+0.042}_{-0.010}$  &$9.3^{+0.8}_{-0.9}$&$10.72^{+0.08}_{-0.07}$&$13.36^{+0.17}_{-0.14}$&$23.8^{+1.7}_{-0.7}$ &$27.3^{+1.2}_{-1.1}$\\
 $\Gamma_{\rm th}$& --- &$4.0^{+0.0}_{-0.6}$&$2.62^{+0.11}_{-0.06}$&$2.258^{+0.011}_{-0.143}$ &$1.80^{+0.05}_{-0.07}$  & $1.400^{+0.016}_{-0.000*}$  \\
   $kT_{\rm e}$(keV) &---& (9)$^\dagger$ & (9)$^\dagger$&$11.1^{+0.9}_{-2.3}$&$9.8^{+0.3}_{-0.7}$&$12.8^{+0.4}_{-0.2}$  \\
$\Gamma_{\rm nth}$ & $2.109^{+0.017}_{-0.009}$&$2.26\pm0.03$&$2.26\pm0.08$&$2.07^{+0.26}_{-0.07*}$&$2.0^{+0.05}_{-0.00*}$ 
&$1.774^{+0.009}_{-0.013}$\\
  $kT_{\rm e}$(keV) & (300)$^\dagger$ & (300)$^\dagger$ & (300)$^\dagger$ & (300)$^\dagger$ & (300)$^\dagger$ & $198^{+102*}_{-90}$\\
 $f_{\rm th}$&(0)$^\dagger$&$0.31^{+0.05}_{-0.03}$&$0.25^{+0.10}_{-0.07}$&$0.67^{+0.07}_{-0.31}$&$0.47^{+0.02}_{-0.08}$ & $0.208^{+0.017}_{-0.019}$\\
\multicolumn{5}{l}{parameters of {\sc kdblur}$^\ddagger$, {\sc xilconv}$^\ddagger$ and Gaussian}\\
$r_{\rm ref,in}(r_{\rm g})$ & $2.2^{+0.5}_{-0.4}$ &$6.8^{+1.3}_{-1.8}$&(10.72)$^\S$&($13.36)^\S$&(23.8)$^\S$ &(27.3)$^\S$ \\
rel\_refl & $2^{+0.0*}_{-0.3}$ &$0.97^{+0.26}_{-0.10}$&$1.0^{+0.00*}_{-0.05}$&$0.86\pm0.08$&$0.69^{+0.07}_{-0.10}$  &$0.68\pm0.07$\\
$E_{\rm c}$(keV)&$6.62^{+0.07}_{-0.08}$ &$6.31^{+0.12}_{-0.11*}$&$6.21^{+0.08}_{-0.00*}$&$6.23^{+0.07}_{-0.03*}$& $6.24^{+0.11}_{-0.04*}$ &$6.20^{+0.12}_{-0.00*}$\\
$\sigma$(keV)& $0.80^{+0.00*}_{-0.06}$& (0.3)$^\dagger$& (0.3)$^\dagger$& (0.3)$^\dagger$& (0.3)$^\dagger$ & (0.3)$^\dagger$\\
norm ($\times 10^{-3}$)& $7.1\pm0.6$ &$16^{+4}_{-5}$&$19\pm 2$&$26\pm3$&$12\pm2$ & $2.1\pm1.1$\\
\hline
EW(eV)& $200\pm17$&$56^{+14}_{-18}$&$82\pm9$&$96\pm11$&$64\pm11$ &$24\pm13$\\
\hline
$\chi^2$(dof) & 156.5(116)& 122.3(115)&103.3(117)&91.4(116)& 93.0(116) & 156.1(115)\\
\hline\hline
$\chi^2$(dof)$^\#$ & 638.7(119) &146.8(117)&276.2(119)&309.6(118)&158.2(118)&166.6(117)\\ 
$F$-value$^\#$($\Delta$dof)  &119.1(3)&11.5(2) & 97.3(2) &138.4(2)&40.7(2)&3.9(2)\\
\hline
  \end{tabular} 
 \label{tab:spec_SSsed-ref}
\end{table*}

To observe the difference between using {\sc smedge} and full reflection continuum, 
we attempted to fit the six sample spectra by replacing {\sc smedge} with the fully ionized reflection code {\sc xilconv}. 
The Gaussian component remained statistically important, similar to the case of the canonical {\sc diskbb} and {\sc nthcomp} fit with reflection in Appendix~\ref{app:diskbb-nthc}.
The resulting {\sc xspec} model is:
\[{\sc tbabs*(SSsed+kdblur*xilconv*SSsed(coronae)+Gaussian)}\]
The parameters used for {\sc xilconv} and {\sc kdblur}  were the same as in Appendix~\ref{app:diskbb-nthc}.
The reflection component was blurred using {\sc kdblur} model which describes the general relativistic blurring with 
$\beta=-3$ and $r_{\rm ref, out}=400 r_{\rm g}$. 
Since the value of $\sigma$ of Gaussian was not well constrained for observations B--F, we fixed it to be 0.3~keV, 
and constrained to be 0.1--0.8~keV for observation A.
Additionally, except for observations A and B, 
the best-fitting value of the inner radius for the reflection $r_{\rm ref, in}$ was consistent with 
$r_{\rm cor}$ of {\sc SSsed} within 90\% uncertainties when it was remained as a free parameter.
Therefore, we tied $r_{\rm ref, in}$ to $r_{\rm cor}$ for observations C--F.
Similar to the approach taken with the phenomenological {\sc smedge} fit,
the inner radius was fixed at $4.5r_{\rm g}$  for observations C--F, 
even though slightly larger $r_{\rm in}$ was estimated for observation B (the SIMS). 
Additionally, $kT_{\rm e}$ of the thermal Comptonising component was fixed at 9~keV for observations B and C. 

\subsection{Summary}
The unabsorbed spectra with the best-fitting reflection continuum (plus Gaussian) are shown in Figs.~\ref{fig:spec_SSsed-ref}a--\ref{fig:spec_SSsed-ref}f 
based on the best-fitting parameters shown in Table~\ref{tab:spec_SSsed-ref}.
The model with the reflection continuum and a Gaussian successfully reproduced the spectra for observations B--E (the SIMS and HIMS). 
As discussed in sections~\ref{sub:soft-sims}--\ref{sub:hims}, the geometry parameters of $r_{\rm in}$ and $r_{\rm cor}$ obtained with full reflections 
were consistent with those approximately obtained by using {\sc smedge}.

In addition, except for the hard state (observation F), the additional Gaussian component remained 
significant (see $F$-values in Table~\ref{tab:spec_SSsed-ref}) and 
had an equivalent width of above 50~eV, indicating contributions from other locations such as winds, or very outer disc as suggested by \cite{connors2020}.
Therefore, the smoothed edge structure that determines the reflection ratio is likely influenced by absorption lines and edges produced by such winds. 
Indeed, XTE~J$1550-564$ is a highly inclined source, and the contribution from winds is expected to be significant.
Complex absorption structures can be clearly observed in the 7--10 keV range, which vary with changes in ionization states.
These complex absorption structures, when combined with the reflection structure, can lead to the possibility of overestimating the values of {\sc rel\_refl}, 
which were estimated to be relatively large as 0.7--1 (Table~\ref{tab:spec_SSsed-ref}).

These issues could potentially be resolved with data sets of finer spectral resolution. However, the analysis performed with our RXTE data has already surpassed
 the intended scope of this paper, and a detailed investigation of absorption/reflection structures is beyond its focus. Instead, we adopted a phenomenological approach using the {\sc smedge} and Gaussian models and estimated the uncertainty without full reflection modelling to be less than 10\% and most likely around 5-8\%.

\subsection{the soft state and SIMS}
The spectrum of observation A was not well reproduced ($\chi^2=156.5$ for degree of freedom of 116) and the model required large reflection continuum 
with {\sc rel\_refl}$=2$ (upper limit).
The SIMS spectrum of observation B was well reproduced by replacing {\sc smedge} with blurred reflection (see Tables~\ref{tab:spec_SSsed} and \ref{tab:spec_SSsed-ref}), and the estimated values of $r_{\rm in}$ and $r_{\rm cor}$ are consistent with those estimated using {\sc smedge}. 
On closer inspection, the best-fitting value of $r_{\rm in}$ increased by 4\% from $(4.4^{+0.4}_{-0.3}) r_{\rm g}$ to $(4.6^{+0.2}_{-0.3})r_{\rm g}$, 
while that of $r_{\rm cor}$ decreased by 3\% from $(9.6^{+1.1}_{-0.9} )r_{\rm g}$ to $(9.3^{+0.8}_{-0.9})r_{\rm g}$.

\subsection{the HIMS}
We again compare the results with {\sc smedge} and with blurred reflection for the HIMS data of observations C, D and E (Table~\ref{tab:spec_SSsed-ref}). 
The values of $r_{\rm cor}$ were found to decrease from 
$(11.9\pm0.3)r_{\rm g}$ to $(10.72^{+0.08}_{-0.07})r_{\rm g}$, from $(15.1^{+0.4}_{-0.6})r_{\rm g}$ to $(13.4^{+0.1}_{-0.2})r_{\rm g}$, 
and from $(25^{+2}_{-1})r_{\rm g}$ to $(23.8^{+1.7}_{-0.7})r_{\rm g}$ for the fixed $r_{\rm in }$ at $4.5r_{\rm g}$, corresponding to decreases of 10\%, 11\%, and 5\%, respectively. 
If we consider 4\% increase in $r_{\rm in}$ (see section~\ref{sub:soft-sims}) and change the fixed value of $r_{\rm in}$ from $4.5r_{\rm g}$ to $4.7r_{\rm g}$, changes in $r_{\rm cor}$ become only 6\%, 8\%, and 1\%, for observations C, D and E, respectively.

\subsection{the bright hard state}
In contrast to the SIMS and HIMS, the goodness of fit significantly deteriorated when we 
replaced {\sc smedge} with the blurred reflection continuum for observation F (Table~\ref{tab:spec_SSsed-ref}).
Additionally, the electron temperature of
the higher temperature Compton component was found to be much higher, exceeding $>100$~keV.
Therefore, a more detailed analyses is required for the two thermal Comptonisations.
 However, the estimated value of $r_{\rm cor}$ was consistent between {\sc smedge} and reflection modellings, and detailed studies to investigate two-temperature Comptonisation is beyond the scope of this paper.

\section{Power Spectral Density of Observations A--F}
\label{app:psd}

\begin{figure*}
\begin{center}
\includegraphics[width=17cm]{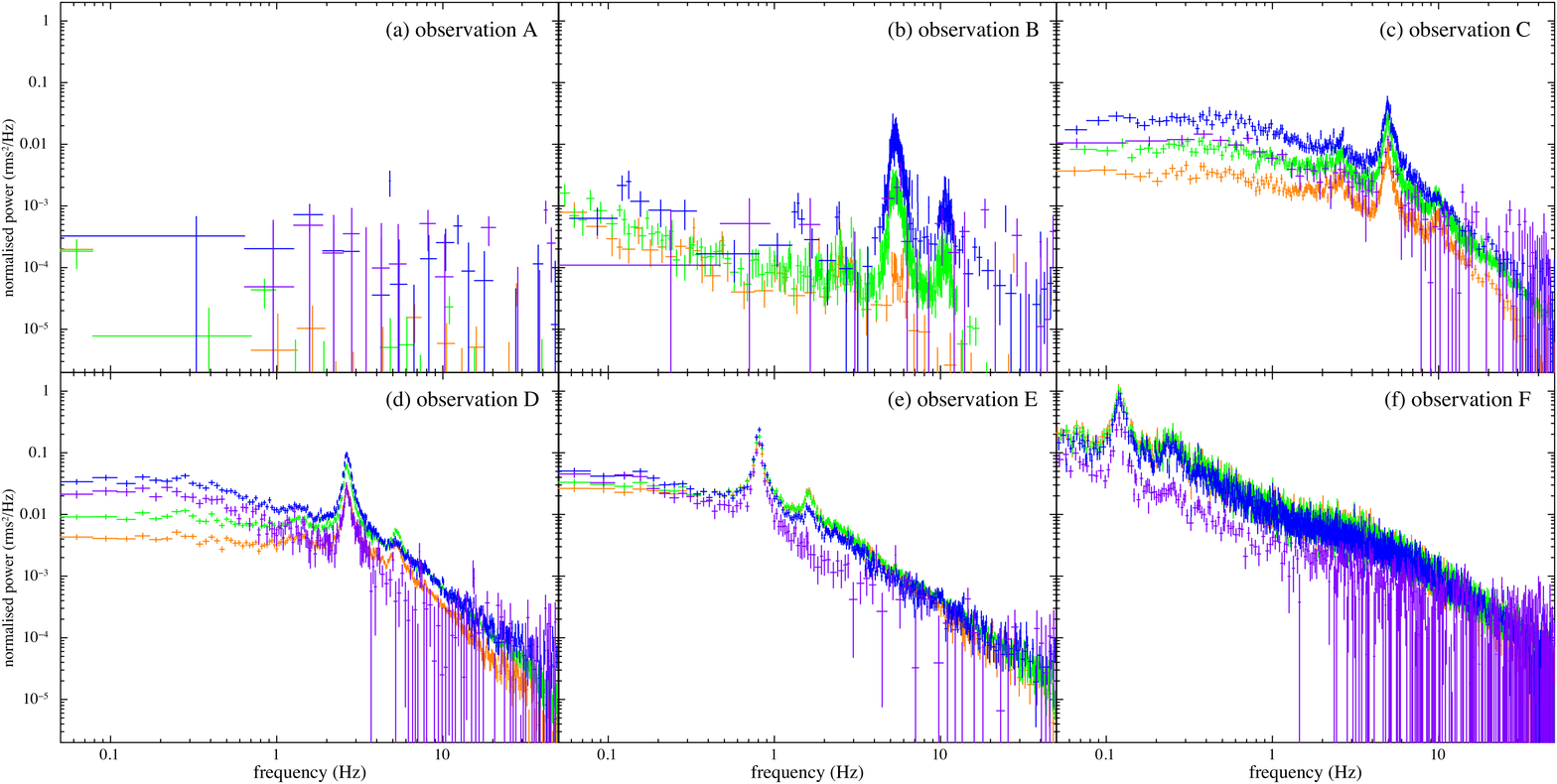}
\end{center}
\caption{Power spectral density normalised to average count rate for observations A--F, 
in the energy bands of 2--4.5~keV (orange), 4.5--15~keV (green), 15--34~keV (blue), and 34--120~keV (purple).
The light curves in the energy range of 2--4.5~keV, 4.5--15~keV, 15--34~keV, and 34--120~keV were extracted from PCA channels of 0--10~ch, 11--35~ch, 36--79~ch, and 80--255~ch, respectively.}
\label{fig:psd}
\end{figure*}

Figure~\ref{fig:psd} displays the PSDs in observations A--F, normalised to the average PCA count rate. In this figure, the PSDs were obtained in different energy bands, including 2--4.5~keV (orange), 4.5--15~keV (green), 
15--34~keV (blue), and 34--120~keV (purple). The PSDs of 2--15~keV and 15--120~keV were obtained using binned mode and event mode, respectively.

These figures demonstrate that the soft state data of observations A does not exhibit any significant time variation, whereas the HIMS and hard state data of observations C--F exhibit significant intensity variations accompanied by strong Type-C QPOs. This is consistent with the association between the QPO types and spectral states shown e.g. by 
\citet{motta2011}.
In the intermediate case of observation B, strong Type-B QPOs are clearly visible, while
the overall variation remains negligible.

\section{The inner radii in the SIMS}
\label{app:rin_sims}

\begin{figure*}
\begin{center}
\includegraphics[width=16cm]{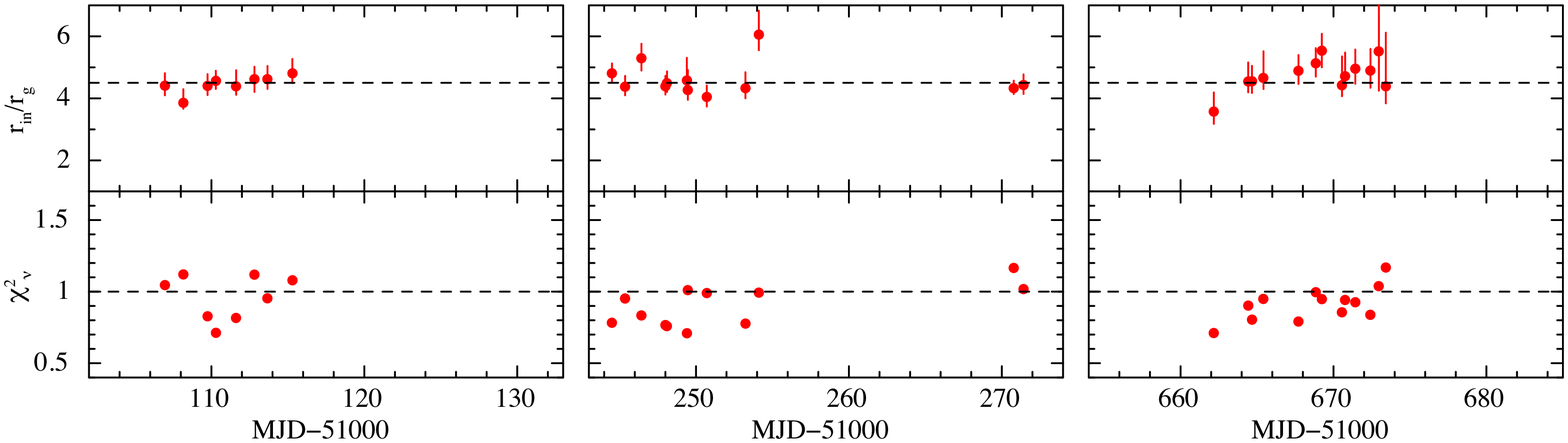}
\end{center}
\caption{Time history of $r_{\rm in}$ in the SIMS . The reduced chi-squared values are plotted in the bottom panels. Dashed lines represent $r_{\rm in}=4.5r_{\rm g}$ (top) and $\chi^2_\nu=1$ (bottom).
Left, middle, and right panels correspond to red, orange and cyan in Fig.~\ref{fig:history_12}.}
\label{fig:rin_sims}
\end{figure*}

Figure~\ref{fig:rin_sims} shows the evolution of the $r_{\rm in}$ values with the {\sc SSsed} model in the SIMS. 
The top panel shows $r_{\rm in}$, which is almost constant at $4.5r_{\rm g}$. 

\end{document}